\documentclass[mnsc,nonblindrev]{informs3}
\OneAndAHalfSpacedXII % current default line spacing
\usepackage{array}

\newcolumntype{L}[1]{>{\raggedright\let\newline\\\arraybackslash\hspace{0pt}}m{#1}}
\newcolumntype{C}[1]{>{\centering\let\newline\\\arraybackslash\hspace{0pt}}m{#1}}
\newcolumntype{R}[1]{>{\raggedleft\let\newline\\\arraybackslash\hspace{0pt}}m{#1}}

\usepackage{natbib}
 \bibpunct[, ]{(}{)}{,}{a}{}{,}%
 \def\bibfont{\small}%
\TheoremsNumberedThrough     % Preferred (Theorem 1, Lemma 1, Theorem 2)
\EquationsNumberedThrough    % Default: (1), (2), ...
\usepackage{natbib}
 \bibpunct[, ]{(}{)}{,}{a}{}{,}%
 \def\bibfont{\small}%

\theoremstyle{example}
\usepackage[hidelinks]{hyperref}
\usepackage{mathabx}
\usepackage{enumitem}
\usepackage{makecell}
\usepackage[ruled,vlined]{algorithm2e}
\usepackage{algpseudocode}
\usepackage{comment}
\usepackage{natbib}
\usepackage{float}
\usepackage{booktabs}

\usepackage{algpseudocode}
 \bibpunct[, ]{(}{)}{,}{a}{}{,}%
 \def\bibfont{\small}%
\makeatletter
\DeclareRobustCommand{\iscircle}{\mathord{\mathpalette\is@circle\relax}}
\newcommand\is@circle[2]{%
  \begingroup
  \sbox\z@{\raisebox{\depth}{$\m@th#1\bigcirc$}}%
  \sbox\tw@{$#1\square$}%
  \resizebox{!}{\ht\tw@}{\usebox{\z@}}%
  \endgroup
}
\makeatother

 \newenvironment{newch}{\color{blue}}{}

\newenvironment{newjhnote}{\color{red} JH: }{}

\usepackage[normalem]{ulem}
\newcommand{\ins}[1]{\textcolor{blue}{{#1}}} % insert
\newcommand{\cmt}[1]{\textcolor{blue}{(CH: {#1})}}

\begin{document}
%%%%%%%%%%%%%%%%

%\setcounter{page}{1}
%\DOI{10.1287/xxxx.0000.0000}%

\RUNAUTHOR{He, Hua, Zhou and Zheng}

\RUNTITLE{Dynamic Portfolio Allocation}

% \TITLE{Bundling Information Goods of Decreasing Value}
\TITLE{Reinforcement-Learning Portfolio Allocation with Dynamic Embedding of Market Information}

\ARTICLEAUTHORS{%
\AUTHOR{Jinghai He}% $^\dagger$}
\AFF{Department of Industrial Engineering \& Operations Research, University of California at Berkeley, Berkeley, CA, 94720, \EMAIL{jinghai\_he@berkeley.edu}}
\AUTHOR{Cheng Hua} % $^\dagger$}
\AFF{Antai College of Economics \& Management, Shanghai Jiao Tong University, Shanghai, China, 200030, \EMAIL{cheng.hua@sjtu.edu.cn}}
\AUTHOR{Chunyang Zhou} % $^\dagger$}
\AFF{Antai College of Economics \& Management, Shanghai Jiao Tong University, Shanghai, China, 200030, \EMAIL{cyzhou@sjtu.edu.cn}}
\AUTHOR{Zeyu Zheng} % \footnote{Corresponding author}}
\AFF{Department of Industrial Engineering \& Operations Research, University of California at Berkeley, Berkeley, CA, 94720, \EMAIL{zyzheng@berkeley.edu}}
} % end of the block

\ABSTRACT{
We develop a portfolio allocation framework that leverages deep learning techniques to address challenges arising from high-dimensional, non-stationary, and low-signal-to-noise market information. Our approach includes a dynamic embedding method that reduces the non-stationary, high-dimensional state space into a lower-dimensional representation. We design a reinforcement learning (RL) framework that integrates generative autoencoders and online meta-learning to dynamically embed market information, enabling the RL agent to focus on the most impactful parts of the state space for portfolio allocation decisions.
Empirical analysis based on the top 500 U.S. stocks demonstrates that our framework outperforms common portfolio benchmarks and the predict-then-optimize (PTO) approach using machine learning, particularly during periods of market stress. Traditional factor models do not fully explain this superior performance. The framework's ability to time volatility reduces its market exposure during turbulent times. Ablation studies confirm the robustness of this performance across various reinforcement learning algorithms. Additionally, the embedding and meta-learning techniques effectively manage the complexities of high-dimensional, noisy, and non-stationary financial data, enhancing both portfolio performance and risk management.
}


% \HISTORY{}

\maketitle
\section{Introduction}\label{sec:introduction}
The pioneering Markowitz portfolio theory \citep{markowitz52portfolio}, a cornerstone of modern investment theory, provides a systematic approach to balancing risk and return in investment decisions. Classical Markowitz portfolio theory typically involve two steps. First, a forecasting model is developed to estimate the distribution of future asset returns. Second, the portfolio weights are determined by optimizing the investor's utility function. This classical Predict-Then-Optimize (PTO) framework has been commonly adopted in the literature. 

However, the complexity and dynamic non-stationarity in the market often pose challenges to the aforementioned classical PTO framework. Firstly, the high-dimensional stochastic nature of stock market data poses challenges for effectively subtracting information from data, in particular, information related to returns and correlations; this point has also been noted in \citep{skyle1993noisemarket, chong2020noisemarket, cong2020alphaportfolio}.
Secondly, the dynamic non-stationary nature of financial markets complicates the task of making accurate predictions over time based on historical data \citep{fama1965investigations, park2011herding, salahuddin2020timevarying}. Many factors related to financial markets can change and evolve rapidly, which not necessarily adhere to the same evolving pattern, including macroeconomic indicators, geopolitical events, and investor sentiment. Traditional statistical and machine learning models often struggle to capture these rapid changes, especially in the long run, leading to outdated predictions that can adversely affect portfolio performance. Thirdly, forecasting errors in the predictive step can be amplified without a clear pattern during the portfolio optimization step, particularly in high-dimensional portfolio optimization settings where the number of assets is large \citep{MI89, ALZ19}.

In this paper, to address the challenges of high-dimensional portfolio allocation in a dynamic non-stationary market, we propose an end-to-end framework named Dynamic Embedding Reinforcement Learning (DERL), which leverages three deep learning methods—deep reinforcement learning, generative encoders, and meta-learning. Firstly, to effectively extract information to interpret stock returns and market dynamics in a high-dimensional environment, we develop a generative encoder to summarize financial market information. The encoder projects high-dimensional raw financial data into lower-dimensional embeddings with more concentrated information, enabling efficient processing of vast amounts of stock market data.
Secondly, we employ online meta-learning to dynamically adjust and adapt the encoder as new data becomes available, forming up-to-date market representations. This allows our framework to automatically update itself to changing and evolving market conditions, capturing non-stationary shifts in market patterns.
Finally, we directly derive the portfolio allocation policy using reinforcement learning. All components in this end-to-end framework ensure that the portfolio allocation adapts to the latest market information, optimizing the investor's utility function in real time.

% Our approach combine three major deep learning methods, namely generative models, meta-learning, and reinforcement learning, and has the following conceptual advantage. Generative models are used to effectively summarize and analyze high-dimensional stock market information, meta-learning is used to adapt the forecasting framework to changing market conditions, and reinforcement learning is used to make portfolio decisions. Together, these methods provide an end-to-end solution for portfolio allocation.
We conduct multiple sets of empirical experiments to validate and explain the performance of the proposed framework with thirty-year data in the U.S. stock market. To ensure the feasibility of trading profits, we follow the suggestions of \cite{avramov2023} and implement certain economic restrictions when constructing the optimal portfolio. First, in the empirical study, we evaluate out-of-sample portfolio performance using top $500$ stocks in terms of market capitalization in each subperiod. Second, to effectively manage portfolio turnover, we follow \cite{Demiguel2020} and incorporate transaction costs into the optimization objective. The Sharpe ratio, a common measure of portfolio performance, is used in this study. The investor is assumed to maximize the Sharpe ratio of net portfolio returns after accounting for transaction costs. Finally, we assume no leverage or short selling is allowed, aligning our strategy with the constraints typically encountered in mutual fund portfolio management.

\subsection*{Empirical Findings}
Empirical results show that our DERL framework achieves significantly higher Sharpe and Sortino ratios compared to the two-step predict-then-optimize (PTO) method using machine learning models, as well as value- and equal-weighted portfolios. We divided the full sample into low and high volatility regimes based on whether the VIX (Volatility Index) published by the CBOE (Chicago Board Options Exchange) is lower or higher than its historical median. The results demonstrate that DERL's outperformance is highly significant under high market volatility conditions compared to low-volatility conditions. This indicates that, compared to other models, the DERL framework is more effective in optimizing investment returns while managing portfolio risk.

Factor analysis shows that the performance of the DERL framework cannot be fully explained by the \cite{FAMA19933} three-factor model or \cite{FAMA19933}-\cite{CAR1997} four-factor model, with the daily risk-adjusted return $\alpha$ exceeding $0.03\%$, or $7.5\%$ per annum. While common factors like momentum and capitalization size are reconstituted monthly or annually, which is less frequent than the daily rebalancing of our DERL portfolio, the estimate of $\alpha$ remains significant across different test periods and volatility regimes. A notable observation is that the DERL framework exhibits timing ability, adjusting its market exposure according to market volatility conditions. Specifically, the portfolio has less market exposure during periods of high volatility compared to periods of low volatility.

We seek to understand the decisions behind the DERL framework by linking the daily stock weights it generates to a set of standard stock characteristics. Using lasso regression on a period-by-period basis, we find those characteristics related to price trends and risks are most frequently chosen by the model. The time-series averages of price-trend coefficients indicate that DERL decisions align with short-term reversal and long-term momentum. Regarding risk characteristics, DERL favors stocks with low systematic risk, which have been volatile over the past $14$ days but have stabilized in the most recent $7$ days. Additionally, DERL demonstrates volatility timing capability, reducing investments in stocks with high systematic risks during periods of market stress.

To elucidate the contributions of the three deep learning methods employed, we conduct a series of ablation exercises and find that the framework's performance remains robust across various reinforcement learning algorithms. Time-series regression analyses reveal that the contribution of the embedding becomes more pronounced when market returns decrease or when the VIX (Volatility Index) increases. This indicates that embedding significantly enhances the model's ability to efficiently process noisy data. Additionally, when market volatility patterns shift, meta-learning boosts model performance by adeptly managing nonstationarity.

\subsection*{Contributions to Literature}
Recently, a significant body of research has applied machine learning (ML) algorithms to predict asset returns and optimize portfolio investments \citep{ban2018machine,KX2023, chen2023deep,jiang2023re}. 
For instance, \cite{gu2020empirical} and \cite{FNW2020} found that using machine learning to integrate large-dimensional firm characteristics improves the predictability of cross-sectional asset returns. They demonstrated that long-short portfolios based on ML-generated signals produce superior out-of-sample performance. \cite{cong2021deep} introduced a deep sequence model for asset pricing, emphasizing its ability to handle high-dimensional, nonlinear, interactive, and dynamic financial data. Their study showed that long-short-term memory (LSTM) with an attention mechanism outperforms conventional models without machine learning in portfolio performance. Additionally, \cite{bryzgalova2023asset} employed an ML-assisted factor analysis approach to estimate latent asset-pricing factors using both cross-sectional and time-series data. Their findings indicate that this method results in higher Sharpe ratios and lower pricing errors compared to conventional approaches when tested on a large-scale set of assets.

% \begin{newjhnote}
% (Need larger scope maybe with more novelty rather than just using another RL agent)
% \end{newjhnote}
We distinguish our study from previous literature in three key aspects. First, the majority of prior studies utilize firm characteristics as model inputs. Although these characteristics exhibit predictive power for future stock returns, they necessitate manual engineering and design for effective prediction. In this paper, our framework inputs only include price-volume information and several technical indicators commonly used by investors. Similar to the convolutional neural network (CNN) approach used by \cite{jiang2023re}, the generative autoencoder in our framework automatically transforms high-dimensional raw inputs into information-concentrated low-dimensional features, significantly reducing the need for manual data selection or transformation. Unlike traditional autoencoders that focus solely on reconstruction, generative autoencoders learn meaningful embeddings to generate realistic new data samples. This results in more robust and informative embeddings that better capture the underlying data distribution.

Second, we incorporate online meta-learning to enable the model to adapt continuously to changing market conditions. Unlike traditional batch learning, which periodically retrains the model using the entire dataset, online meta-learning updates the model incrementally. As new data points are received, the model can quickly adjust its parameters without requiring a complete retraining process, significantly reducing computational intensity. This is particularly advantageous given that batch retraining of ML models is relatively infrequent due to the intensive computation required (see, e.g., \cite{gu2020empirical} and \cite{cong2020alphaportfolio}). By using online meta-learning, our model can continuously learn and adapt, making it well-suited for the dynamic nature of financial markets.

Finally, we propose an end-to-end reinforcement learning (RL) framework that automatically and directly provides daily weights for each asset as outputs. RL is an emerging branch of statistical and machine learning algorithms, and its application in portfolio allocation is still evolving. In a pioneering work, \cite{cong2020alphaportfolio} first applied policy-based RL to solve the dynamic portfolio allocation problem with high-dimensional state variables, demonstrating superior performance. Unlike their approach, which computes a score and selects the top and bottom $d$ equities based on that score, our framework directly outputs the allocation percentage for each equity in the portfolio. Additionally, while \cite{cong2020alphaportfolio} use firm characteristics as inputs and conduct monthly adjustments, our method relies on daily adjustments solely based on price-volume data and technical indicators. Our comprehensive framework incorporates dynamic market embedding and demonstrates robustness across various state-of-the-art RL algorithms. 
Complementing their study, we demonstrate the superior performance of end-to-end strategies compared to the traditional two-step framework.

\begin{comment}

{\newch More details about the empirical results shall be presented 

Empirically, our framework outperforms traditional approaches and provides an insight into the end-to-end trading process and the ability to dynamically capture changes in the financial market.
}

{\newch Finance market dynamics}
\end{comment}

Our paper is organized according to the following structure. In \S \ref{sec:methodology}, we set up the model and present our methodology. In \S \ref{sec:numerical} we present our empirical studies using U.S. equities.  In \S \ref{sec: conclusion}, we summarize our results and the corresponding managerial insights into portfolio management and algorithmic trading. We present more implementation details of our algorithms and detailed disccusions of related literature in the E-Companion.

\section{Methodology}
\label{sec:methodology}
%\subsection{Overview and Notation}
In this section, we first present a generic reinforcement learning framework for portfolio allocation that can incorporate diverse types of market information inputs in \S \ref{sec: Main framework}. Next, we describe the generative encoder used to encode raw market information into low-dimensional embeddings in \S \ref{sec:generative models and autoencoders}. We then explain how these embeddings are dynamically updated using online meta-learning. Finally, we integrate all three components to introduce our Dynamic Embedding Reinforcement Learning (DERL) framework in \S \ref{sec: main of Embed then RL}.

\subsection{Portfolio Allocation via Reinforcement Learning}\label{sec: Main framework}

%In this section, we present a reinforcement learning (RL) formulation for dynamic portfolio allocation of stocks through buy-sell decisions with the aim of maximizing expected accumulated discounted revenue over a specified number of epochs $T$. No leverage or short selling is allowed so that no exploitative use of them can happen. For simplicity, we assume no market impact and therefore do not incorporate market impact models. Our framework models the stock market as a system in which public market information and the agent's private asset levels are considered as states $(\boldsymbol{s})$, and buy-sell decisions are treated as actions $(\boldsymbol{a})$. We assume a market with $D$ different tradable stocks and no leverage or short selling, which aligns more closely with situations typically encountered in mutual fund portfolio management. 
%We model the trading process as a Markov decision process (MDP). We describe the market information and our asset levels as states and buy-sell decisions as actions in the MDP. We consider a market of $D$ different tradable stocks. For simplicity consideration, leverage and short selling are not allowed in our model, which corresponds more to situations in mutual funds rather than in hedge funds. 

%In this section, we present a reinforcement learning (RL) formulation for dynamic portfolio allocation. 
We consider an investor aiming to optimize portfolio performance over the next $T$ periods by investing in $D$ different assets (including equities and a risk-free asset). 
% No leverage or short selling is allowed which aligns more closely with situations typically encountered in mutual fund portfolio management. For simplicity, we assume no market impact and therefore do not incorporate market impact models. 
Our framework models the equity market as a system where public market information and current holding positions are considered states \((\boldsymbol{s})\), and the weights of equities and the risk-free asset in the portfolio at each decision step are treated as actions \((\boldsymbol{a})\). The investor makes portfolio decisions based on the state at each step to maximize utility, specifically the portfolio performance over the following \(T\) periods. 

In this study, we focus on daily end-of-day trading, where the investor makes a single trading decision for all equities each day, with trading orders executed based on the closing prices of equities at the end of each trading day. Our framework relies solely on price and volume information for decision-making, similar to \cite{jiang2023re}, and uses the Sharpe ratio as the measure of the investor's utility, as in \cite{cong2020alphaportfolio}. Notably, our framework is flexible and can accommodate various types of input, such as stock characteristics, news, and macroeconomic information. Additionally, it can be adapted to other trading strategies or utility functions.

\subsubsection{Formulation of Reinforcement Learning}
Reinforcement learning (RL) comprises a set of algorithms designed to train an intelligent agent to make autonomous decisions through interaction with an environment. This interaction is typically modeled as a Markov decision process, denoted as \( M = \{\mathcal{S}, \mathcal{A}, \mathbb{P}, r, \gamma\} \). In this model, \(\mathcal{S}\) represents the set of possible states within the environment, \(\mathcal{A}\) denotes the set of feasible actions that the agent can take, \(\mathbb{P}\) characterizes the state transition probabilities influenced by the agent's actions, \(r\) signifies a scalar reward obtained from taking specific actions in given states, and \(\gamma\) is the discount factor determining the importance of future rewards, similar to the discount rate used for valuing cash flows.
In the remainder of this section, we introduce the modeling of portfolio allocation in an RL setting.

The market state \(\boldsymbol{s} = (\boldsymbol{\delta}^\top, \boldsymbol{w}^\top, \boldsymbol{l}^\top, x)^\top \in \mathcal{S} \subseteq \mathbb{R}^{2D + h + 1}\) is a collection of market information that affects portfolio decisions. It includes the \(D\) assets' returns \(\boldsymbol{\delta} \in \mathbb{R}^{D}\), weights of current equity and risk-free asset holdings \(\boldsymbol{w} \in \mathbb{R}_{0}^{D+}\),  market-metrics \(\boldsymbol{l} \in \mathbb{R}^{h}\) that captures information including price-volume information, technical indicators, news and macroeconomic information, and total current wealth \(x \in \mathbb{R}_{0}^{+}\)\footnote{For cash (risk-free) asset, its price is always $1$ and return is the risk-free interest rate.}. Specifically for \(\boldsymbol{l}\), in this work, we only consider price-volume information and technical indicators for the equities, although it can also incorporate other relevant market information, including stock characteristics, fundamental information, and macroeconomic information.

\begin{comment}
    and is generally divided into two categories. The first category includes variables useful for explaining equity returns and thereby improving portfolio performance. To reduce the burden of feature engineering, extract predictive information from raw market trading data and commonly used technical indicators. The second category consists of the historical weights of each equity in the portfolio, which directly affect upcoming portfolio decisions by influencing transaction costs.
\end{comment}

The action $\boldsymbol{a} \in \mathcal{A}\subseteq \mathbb{R}^{D}$ is a vector of asset weights, where the $d$-th entry $a^{[d]}$ represents the weight of asset $d$ in the portfolio, and $\mathcal{A}$ is the set of feasible actions. In this work, no leverage or short selling is allowed, which aligns with typical mutual fund portfolio management practices. Under the no short-selling constraint, the equity weights satisfy $\sum_{d=1}^{D}{a^{[d]}}=1$ and $a^{[d]} \geq 0$ for $d=1,\cdots, D$, including the risk-free asset\footnote{To ensure the constraint is satisfied, we can apply the softmax operation after the final layer. The softmax function normalizes the actions so they sum to 1 and ensures each action is between 0 and 1, which follows $a^{[d]}=e^{a^{[d]}}/(\sum_{i=1}^D e^{a^{[i]}})\in[0,1]$ and $\sum_{d=1}^D a^{[d]} = 1$. Our setting can also be adapted to the long-short setting. For long-short settings, we only need the constraint that the weight actions sum to 1. In this case, we can apply the following transformation: $a^{[d]}\leftarrow a^{[d]}-\frac{1}{D}\left(\sum_{i=1}^D a^{[i]}-1\right),\forall a^{[d]}\in \mathbb{R}$.}. One key connection between action and state is that the action $\boldsymbol{a}_t$ taken at time $t$ will be the asset weight information $\boldsymbol{w}_{t+1}$ at time $t+1$, i.e., $\boldsymbol{w}_{t+1} = \boldsymbol{a}_{t}$. 

The transition probability $\mathbb{P}(\boldsymbol{s}^{'} |\boldsymbol{s}, \boldsymbol{a})$ represents the probability of transitioning to a new market state $\boldsymbol{s}'$ when taking action $\boldsymbol{a}$ in the current state $\boldsymbol{s}$. The stochasticity of the transition dynamics stems from the uncertainty surrounding the return vector $\boldsymbol{\delta'}$ and market-metrics $\boldsymbol{l}'$ on the next day. Once the next day arrives and the return $\boldsymbol{\delta'}$ and auxiliary information $\boldsymbol{l}'$ are revealed, we can calculate the components in $\boldsymbol{s}'$ as follows
\begin{equation}
\begin{aligned}
    \boldsymbol{w}' =\boldsymbol{a}, \quad
    x' = \boldsymbol{\delta}'^\top\boldsymbol{w}\cdot x -c(\boldsymbol{a}, \boldsymbol{w}),    
\end{aligned}
\end{equation}
where $c(\boldsymbol{a}, \boldsymbol{w})$ denotes the transaction cost of executing the action $\boldsymbol{a}$ when the current holding is $\boldsymbol{w}$, which includes factors such as commissions and spreads. 

After taking action $\boldsymbol{a}_{t}$ in the $t$-th step, the agent receives an instant return on the whole portfolio $R_t = \frac{x_{t+1}-x_t}{x_t}$. To capture the utility of the investor and the long-term effect of the actions, similar to \cite{cong2020alphaportfolio}, we use the Sharpe ratio to measure portfolio performance, which serves as the final reward for the reinforcement learning agent. We have
\begin{equation}
    r_t = \frac{\mu_t}{\sigma_t},
\end{equation}
where $\mu_t = \frac{1}{k}\sum_{i=t}^{t+k-1} R_i$ and $\sigma_t = \sqrt{\frac{1}{k-1}\sum_{i=t}^{t+k-1} (R_i-\mu_t)^2}$ are the mean and standard deviation of the realized portfolio return in the following $k$ days after taking action $\boldsymbol{a}_t$, respectively, in excess of the risk-free rate and net of transaction costs.

\subsubsection{The Objective of Reinforcement Learning}
The objective of RL for portfolio allocation is to learn
a trading policy that maximizes the expected long-term (discounted) value of the portfolio. 

Formally, a trading policy is represented as  $\pi(\boldsymbol{a}|\boldsymbol{s})\in \Pi:\mathcal{S}\times\mathcal{A}\rightarrow \Delta(\mathcal{A})$, specifying the probability distribution over the set of actions $\mathcal{A}$ when in state $\boldsymbol{s}$. Here, $\Delta(\mathcal{A})$ denotes the simplex of probability distributions over the action space. Given a fixed policy $\pi$, the state transition dynamics can be determined as follows:
\begin{equation}
    \mathbb{P}^\pi(\boldsymbol{s}'|\boldsymbol{s}) = \int_{\boldsymbol{a}\in\mathcal{A}(\boldsymbol{s})}\pi(\boldsymbol{a}|\boldsymbol{s}) \mathbb{P}(\boldsymbol{s}'|\boldsymbol{s},\boldsymbol{a}) \mathrm{d}  \boldsymbol{a}.
\end{equation}
With the state transition dynamics $\mathbb{P}^\pi(\boldsymbol{s}'|\boldsymbol{s})$, we can calculate the probability of any trajectory $\boldsymbol{\tau}^\pi(\boldsymbol{s}_0,\boldsymbol{a}_0,\boldsymbol{s}_1\cdots,\boldsymbol{s_T})$. By taking the expectation over all trajectories, we can estimate the expected sum of discounted future returns. We define the value function $V^{\pi}_t(\boldsymbol{s}): \Pi\times S\times [T]\rightarrow\mathbb{R}$ as the expected cumulative discounted return when visiting state $\boldsymbol{s}$ at time $t\leq T$:
\begin{equation}
V^{\pi}_t(\boldsymbol{s})=\mathbb{E}_{\boldsymbol{\tau}^\pi}\left[\sum_{k=t}^{T} \gamma^{k-t} r_{k}  \mid \boldsymbol{s}_{t}=\boldsymbol{s}\right].
\end{equation}

The aim of reinforcement learning (RL) is to find the optimal policy $\pi^\star(\boldsymbol{a}|\boldsymbol{s})$ that maximizes the expected value function for any $\boldsymbol{s}$. This indicates that $\forall s\in \mathcal{S}$, we have  
\begin{equation}
\pi^{\star}=\arg \max _{\pi \in \Pi} V^{\pi}(\boldsymbol{s}).
\end{equation}

In modern RL practice, researchers typically approximate the value function directly when the dimensionality of states or actions is high, rather than attempting to estimate the transition dynamics $\mathbb{P}(\boldsymbol{s}'|\boldsymbol{s},\boldsymbol{a})$. This value function approximation approach forms the basis of  \textit{model-free} RL algorithms \citep{silver2014ddpg, silver2016mastering, fujimoto2018td3}. These algorithms use various function types (like neural networks) and techniques to approximate the value function induced by a given policy. For more details on model-free RL with value function approximation, readers can refer to \S \ref{ec:details of RL}.

Our framework employs model-free RL agents due to the difficulty of directly modeling the transition dynamics in financial markets. However, applying model-free reinforcement learning in dynamic portfolio allocation remains challenging due to the large number of assets, high-dimensional factors associated with each asset, and the excessive random noise present in high-dimensional financial data \citep{dynamic_datasets}. To address these challenges, we propose developing embeddings for the high-dimensional state space as inputs to our reinforcement learning framework. In the following section, we discuss how to develop effective and efficient stock market embeddings using a generative autoencoder.

\subsection{Generative Autoencoder for State Embedding}\label{sec:generative models and autoencoders} 

To address the challenges posed by high dimensionality and low signal-to-noise ratio in financial data, we use embeddings, which are lower-dimensional representations of the original high-dimensional space that retain relevant information and facilitate the learning of features. By reducing noise and redundant information, embeddings enhance a model's ability to generalize, making it easier to extract meaningful patterns and relationships. Additionally, embeddings can incorporate extra information, such as transition dynamics, that may be difficult to capture in raw data. By encoding this information in the embedding space, the model can make more informed decisions and better handle the complexities of financial data.
% Furthermore, embeddings increase the signal-to-noise ratio by minimizing random noise and highlighting the underlying structure of the data. This could lead to more accurate predictions and better investment decisions.

In this paper, we use generative autoencoders to embed original states into low-dimensional representations, enabling the reinforcement learning (RL) agent to process these inputs more efficiently. Unlike previous encoders, such as DynE \citep{whitney2019dynamics} and autoencoders for asset pricing \citep{gu2021autoencoder}, which directly map information into embeddings based on state distance, our framework learns a mapping that embeds states and actions while incorporating market transition information. This approach ensures that nearby embeddings have similar distributions for the next state, allowing the RL agent to make more informed decisions by effectively capturing the dynamics of financial markets.

\subsubsection{Generative Autoencoders}
% Autoencoders are widely used in dimensionality reduction and feature extraction, and have been successfully in applications such as image compression, denoising and generation \citep{chang2021mage,tolstikhin2018wasserstein}, drug discovery \citep{polykovskiy2018application}, and language models such as ChatGPT. 

Autoencoders are a type of neural network used for unsupervised learning that aim to learn a compressed representation (embedding) of input data and then reconstruct the data from this embedding. Generative autoencoders extend the concept of autoencoders by enforcing a structured latent space and focusing on the underlying data distribution, providing more robust and informative embeddings compared to regular autoencoders.

% \ins{(Formally, the encodings $\boldsymbol{z_{s}}$ are optimized to form a compressed representation of the sufficient statistics of $\mathbb{P}\left(\boldsymbol{s^{\prime}} \mid \boldsymbol{s}, \boldsymbol{a}\right)$ such that $\mathbb{P}\left(\boldsymbol{s^{\prime}} \mid \boldsymbol{s}, \boldsymbol{a}\right) \approx \mathbb{P}\left(\boldsymbol{z_s^{\prime}} \mid \boldsymbol{z_{s}}, \boldsymbol{a} \right)$. )}

Formally, generative autoencoders are a set of probabilistic models that learn a continuous and low-dimension embedding $\boldsymbol{z}\in \mathcal{Z}\subseteq \mathbb{R}^{\dim(\mathcal{Z})}$ (also called a latent variable) for the original variable $\boldsymbol{s}\in \mathcal{S}\subseteq \mathbb{R}^{\dim(\mathcal{S})}$. Generative autoencoders are designed to learn a representative embedding that can reconstruct the original data. The learnt embedding can further be used generate new data. Typically, the dimension of the embedding is substantially smaller than the dimension of the original input, i.e., $\dim(\mathcal{Z})\ll \dim(\mathcal{S})$. A generative autoencoder includes:
\begin{itemize}
    \item an encoder $\Gamma_\phi(\boldsymbol{z}|\boldsymbol{s})$ with parameters $\phi$, which maps each $\boldsymbol{s}$ to a distribution on the latent variable $\boldsymbol{z}$; 
    \item a decoder $G_\theta(\boldsymbol{s}|\boldsymbol{z})$ with parameters  $\theta$, which maps $\boldsymbol{z}$ to a distribution over the original variable $\boldsymbol{s}$.   
\end{itemize}
During training, these two components work sequentially. The encoder first maps the raw variable $\boldsymbol{s}$ to a latent variable $\boldsymbol{z}$, and then the decoder reconstructs the original variable from the latent representation. This process can be interpreted as \textit{encoding} the information in the raw variable into a lower-dimensional latent space and then \textit{decoding} it back to the original space, i.e., 
\begin{equation}\label{eq:flow}
    \boldsymbol{s} \xrightarrow[\text{encode}]{\Gamma_\phi} \boldsymbol{z}(\boldsymbol{s}) \xrightarrow[\text{decode}]{G_\theta} \boldsymbol{\boldsymbol{s}}.
\end{equation}

A well-trained generative autoencoder can work separately with its two components. Using the encoder, high-dimensional and noisy input $\boldsymbol{s}$ can be compressed into a low-dimensional representation $\boldsymbol{z}(\boldsymbol{s})$ (i.e. $\boldsymbol{s} \rightarrow \boldsymbol{z}$). This $\boldsymbol{z(s)}$ is usually more information-concentrated, computationally efficient, and can capture valuable information for specific downstream tasks. Similarly, with the decoder, we can generate $\boldsymbol{s}$ for any $\boldsymbol{z}$ (i.e. $\boldsymbol{z} \rightarrow \boldsymbol{s}(\boldsymbol{z})$). 

We present the details, some theoretical properties of generative auto-encoders and different types of autoencoders that can fit into our framework in \S \ref{ec:autoencoders}.
% We include complete details about the generative autoencoder in the E-Companion.

%We use a probabilistic encoder on the state space, based on simulated path training, which maps high-dimensional states to a lower-dimensional latent variable $Z$ to capture the full information of transition dynamics while remaining computationally efficient, allowing the RL agent to make use of all market information.
\subsubsection{State Embedding}
Different from conventional use of generative encoders that aim to regenerate the data itself, we use generative autoencoders to capture hidden transition factors in our RL-based portfolio management framework. 
Recall that in the RL setting, $\boldsymbol{s}\in \mathcal{S}$ represents the current state, $\boldsymbol{a}\in \mathcal{A}(\boldsymbol{s})$ represents the current action,  $\boldsymbol{s^{\prime}}\in \mathcal{S}$ represents the next state. We introduce the embedded variable $\boldsymbol{\boldsymbol{z_s}}\in \mathcal{Z}$ for state $\boldsymbol{s}$. Our goal is to train a generative autoencoder whose encoder $\Gamma_\phi$ can provide a summarized and low-noise-contained embedding $\boldsymbol{z_s}\in\mathcal{Z}$ for state $\boldsymbol{s}$. Instead of only allowing  $\boldsymbol{z_s}$ to contain sufficient information to reconstruct $\boldsymbol{s}$ in Equation \eqref{eq:flow}, we aim to find $\boldsymbol{z_s}$ that can reveal transition information. Therefore, we focus on finding the latent representation $\boldsymbol{z_s}$ that can reconstruct the next state $\boldsymbol{s^\prime}$, given $\boldsymbol{a}\in\mathcal{A}(\boldsymbol{s})$:
\begin{equation}\label{eq:transition equation}
\boldsymbol{s} \xrightarrow[\text{encode}]{\Gamma_\phi} \boldsymbol{\boldsymbol{z_s}} \xrightarrow[\text{{decode with }} \boldsymbol{a}\in\mathcal{A}(s)]{G_\theta} \boldsymbol{s^\prime}
\end{equation}

\begin{figure}
    \FIGURE
    {\includegraphics[scale=0.6]{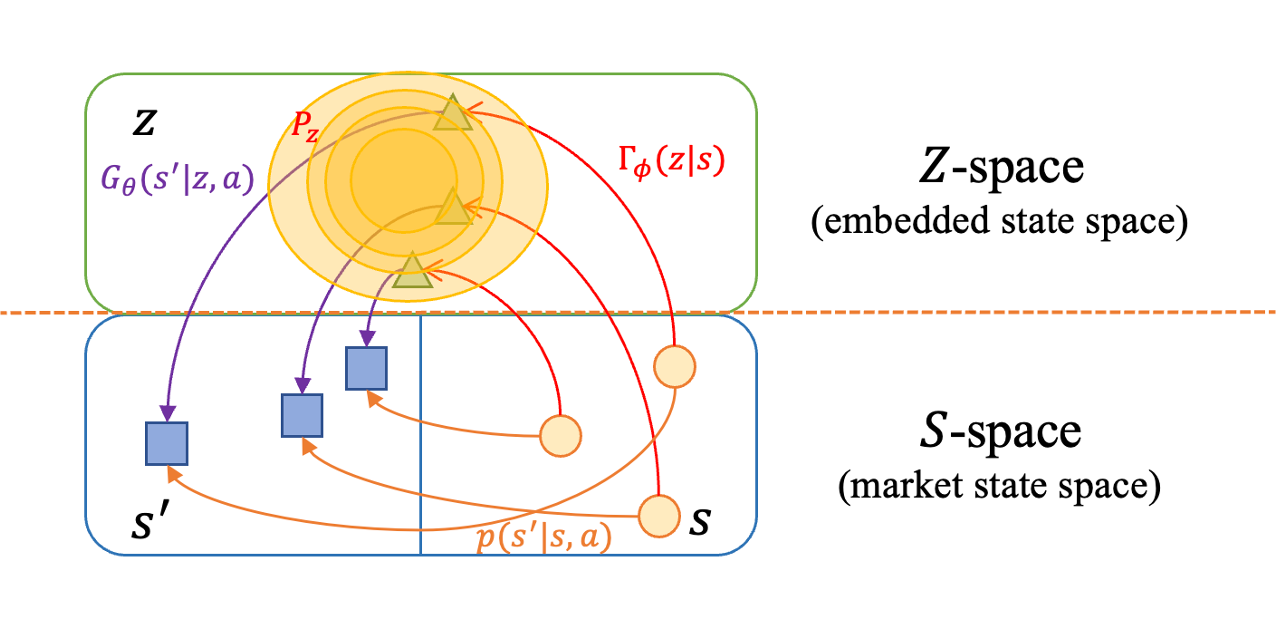}}
    {State Embedding with Generative Autoencoders. \label{fig:generative models}}
    {The upper half of the figure represents the latent space $\mathcal{Z}$ with lower dimensionality. The lower half represents the original state space of the financial market, including current states $\boldsymbol{s}$ ($\circ$) and next states $\boldsymbol{s^\prime}$ ($\square$). We aim to train a generative autoencoder where the encoded states $\boldsymbol{z}$ from $\Gamma_\phi$ are used by the decoder $G_\theta$ to generate states based on a given action $\boldsymbol{a}$, matching the true next states $\boldsymbol{s^{\prime}}$. The embedding $\boldsymbol{z}$ provides a low-dimensional representation of the original market state.}
\end{figure}

Figure \ref{fig:generative models} illustrates how we use generative autoencoders to find a latent state embedding $\boldsymbol{\boldsymbol{z_s}}$ that captures transition dynamics.
The intuition behind the embedding $\boldsymbol{z}_s$ is that it allows us to decompose the transition dynamics $\mathbb{P}(\boldsymbol{s}'|\boldsymbol{a},\boldsymbol{s})$ into
\begin{equation}\label{eq:decompose}
\mathbb{P}(\boldsymbol{s}'|\boldsymbol{a},\boldsymbol{s})=\int_{\boldsymbol{z_s}\in\mathcal{Z}}\Gamma_\phi(\boldsymbol{z_s}|\boldsymbol{s})G_\theta(\boldsymbol{s}'|\boldsymbol{z_s},\boldsymbol{a}) \text{d}\boldsymbol{z_s},
\end{equation}
where $\Gamma_\phi(\boldsymbol{z_s}|\boldsymbol{s})$ is the encoder that maps the raw state $\boldsymbol{s}$ to the embedded state $\boldsymbol{z_s}$, and $G_\theta(\boldsymbol{s}'|\boldsymbol{z_s},\boldsymbol{a})$ is the decoder that generates the next state from the embedded state and action. This decomposition is important because it allows the model to break down the complex transition dynamics into more manageable components, facilitating learning and representation of state transitions in reinforcement learning tasks.

In generative autoencoders, the encoder $\Gamma_\phi(\boldsymbol{z_s}|\boldsymbol{s})$ is typically probabilistic, meaning it defines a distribution over $\boldsymbol{z_s}$. This probabilistic nature is useful in our portfolio allocation problem because it provides a more robust representation of market states, accounting for uncertainty and variability. Besides, the embedding $\boldsymbol{z_s}$ has  more concentrated information and higher signal-to-noise (SNR) ratio than the original state $s$, considering it summarizes information for constructing next state with significantly lower dimension.
In our framework, we only need the encoder $\Gamma_\phi(\boldsymbol{z_s}|\boldsymbol{s})$ in a trained autoencoders, as it provide the downstream RL task with informative and low-dimensional representation of the raw market states.
% embedded state.

\subsubsection{Training Generative Autoencoders for State Embeddings}
The training process of our generative autoencoder involves finding the encoder $\Gamma_\phi$ and the decoder $G_\theta$ that minimize the expected distance between the true next state and the reconstructed next state for all possible tuples $(\boldsymbol{s},\boldsymbol{a},\boldsymbol{s}')$, given by
\begin{equation} \label{eq: generative loss}
    \min_{\phi,\theta} \mathcal{L}(\phi,\theta) = \min_{\phi,\theta} \mathbb{E}\left[\mathcal{C}\left(\boldsymbol{s}',\mathbb{E}_{G_\theta(\hat{s}'|z_s\sim\Gamma_\phi(z_s|s),a)}\left[\boldsymbol{\hat{s}}'\right]\right) \right],
\end{equation}
where $\mathcal{L}(\phi,\theta)$ represents the loss function, $\mathbb{E}_{G_\theta(s'|z_s\sim\Gamma(z_s|s),a)}\left[\boldsymbol{\hat{s}}'\right]$ is the expected reconstructed next state, and $\mathcal{C}:\mathcal{S}\times\mathcal{S}\rightarrow \mathbb{R}$ is a distance metric that measures the dissimilarity between the reconstructed next state $\boldsymbol{\hat{s}}'$ and the true next state $\boldsymbol{s}'$.  The steps to construct and apply the loss function \eqref{eq: generative loss} are as follows:
\begin{itemize}
    \item For each state $\boldsymbol{s}$, sample $\boldsymbol{z}_s$ from the current encoder $\boldsymbol{z}_s\sim\Gamma_\phi(\cdot|\boldsymbol{s})$;
    \item Take a random action $\boldsymbol{a}\in\mathcal{A}(\boldsymbol{s})$, and compute the expected next state $\hat{\boldsymbol{s}}'$ using the decoder distribution $G_\theta(\cdot|\boldsymbol{z_s},a)$;
    \item  Measure the dissimilarity between the true next state $\boldsymbol{s}'\sim\mathbb{P}(\boldsymbol{s}'|\boldsymbol{s},\boldsymbol{a})$ and the reconstructed state $\boldsymbol{\hat{s}}'$ using the distance metric $\mathcal{C}(\boldsymbol{s}',\hat{\boldsymbol{s}}')$, and update the parameters $\theta,\phi$ using a gradient-based method.
\end{itemize}

To obtain the embedding, various generative autoencoder structures can be used, such as the Variational Autoencoder (VAE) \citep{Kingma2014AutoEncodingVB}, Adversarial Variational Bayes Autoencoders \citep{mescheder2017adversarial}, and Wasserstein Autoencoder \citep{tolstikhin2018wasserstein}. These autoencoders differ mainly in their distance metrics $\mathcal{C}$ and sampling rules in the first two steps. 
% However, once the autoencoder is well-trained, regardless of the variant used, the $\boldsymbol{z}_s$ derived from $\Gamma_\phi(\cdot|\boldsymbol{s})$ can capture transition information of $\boldsymbol{s}$ for all $\boldsymbol{a}\in\mathcal{A}(\boldsymbol{s})$. In this paper, we explore the use of variational autoencoder (VAE) and Wasserstein autoencoder (WAE) in our framework, as discussed in \S \ref{sec: main of Embed then RL}.

Once the autoencoder is trained, we replace all states \(\boldsymbol{s}\) in the RL setting mentioned in \S \ref{sec: Main framework} with their corresponding embeddings \(\boldsymbol{z_s}\). In other words, the RL agent generates a policy \(\pi(\boldsymbol{a}|\boldsymbol{z}_s)\) based on the embedded states. This embedding reduces the computational complexity of the RL algorithm and enhances the stability of the learning process due to its low-dimensional and high signal-to-noise nature. One potential limitation of this embedding is that it is trained with historical data, and if the dynamics captured in Equation (\ref{eq:decompose}) change, the trained embedding may fail to account for nonstationarity in the market transitions. Therefore, it may be necessary to develop an approach to incorporate new market transition information over time.

\begin{comment}
Instead of considering modeling the transition probability $\mathbb{P}(\boldsymbol{s}'|\boldsymbol{s},\boldsymbol{a})$, we further model the joint probability distribution of the four random variables $\mathbb{P}(\boldsymbol{s}, \boldsymbol{s^{\prime}}, \boldsymbol{z_s},\boldsymbol{a}):$ $\mathcal{S} \times \mathcal{S} \times \mathcal{Z} \times \mathcal{A}(\boldsymbol{s}) \rightarrow [0,1]$. We also denote by $\mathbb{P}_{G, Z}(\boldsymbol{s^{\prime}}, \boldsymbol{z_s}, \boldsymbol{a}) = \mathbb{E}_{\boldsymbol{s}\in \mathcal{S}}\left[ \mathbb{P}(\boldsymbol{s}, \boldsymbol{s^{\prime}}, \boldsymbol{z_s}, \boldsymbol{a})\right]$ the joint distribution of variables $(\boldsymbol{s^{\prime}}, \boldsymbol{z_s}, \boldsymbol{a})$,
%, where $Z$ is first sampled from embed $P_{Z}$ and next $Y$ from $P_{G}(Y \mid Z, A)$. 
and $\mathbb{P}_{G}$ the marginal distribution of $\boldsymbol{s^\prime}$ when $(\boldsymbol{s^\prime}, \boldsymbol{z_s}, \boldsymbol{a}) \sim \mathbb{P}_{G, Z}$, which indicates $\mathbb{E}_{\boldsymbol{a}\in \mathcal{A}(s),\boldsymbol{z_s}\in\mathcal{Z}}\left[ \mathbb{P}_{G,Z}(\boldsymbol{s^{\prime}}, \boldsymbol{z_s}, \boldsymbol{a})\right]$.
\end{comment}

\subsection{Dynamic Embedding Update Using Meta-Learning} 
\label{sec:fine-tuning}
Market components, such as return patterns \citep{salahuddin2020timevarying}, price series \citep{fama1965investigations}, and risk loadings \citep{sunder1980stationarity}, change over time. A static model will not capture sufficient market information. Conventional methods require model retraining at intervals. However, due to the intensive computation required, batch retraining of the ML model is relatively infrequent (e.g., \citealt{gu2020empirical}). This can lead to poor performance when the market shifts, and the model fails to capture key dynamics. To address this, our framework dynamically updates the encoder over time to quickly adapt to new market dynamics. We incorporate online meta-learning techniques, inspired by \cite{rajasegaran2022fully}. Unlike traditional batch learning, online meta-learning updates the model incrementally. As new data points are received, the model can quickly adjust its parameters without the need for complete retraining. This approach significantly reduces computational intensity compared to batch learning while effectively capturing market changes.

% In addition to the encoder and RL agent trained based on historical data, we also consider that the market may change over time \citep{salahuddin2020timevarying,yuan2011exchange}, in which case the embedding will not capture sufficient market information if it does not change with it. Therefore, our strategy is to update the encoder as time changes. To learn a more general representation that can quickly adapt to new market dynamics, we update the encoder through a fully online meta-learning (FOML) framework \citep{rajasegaran2022fully}. This meta-learning framework updates the encoder with the most recent acquired information in an online learning fashion, as shown in Figure \ref{fig:dynamically update}.

\begin{figure}[htbp]
    \FIGURE
    {\includegraphics[scale=0.7]{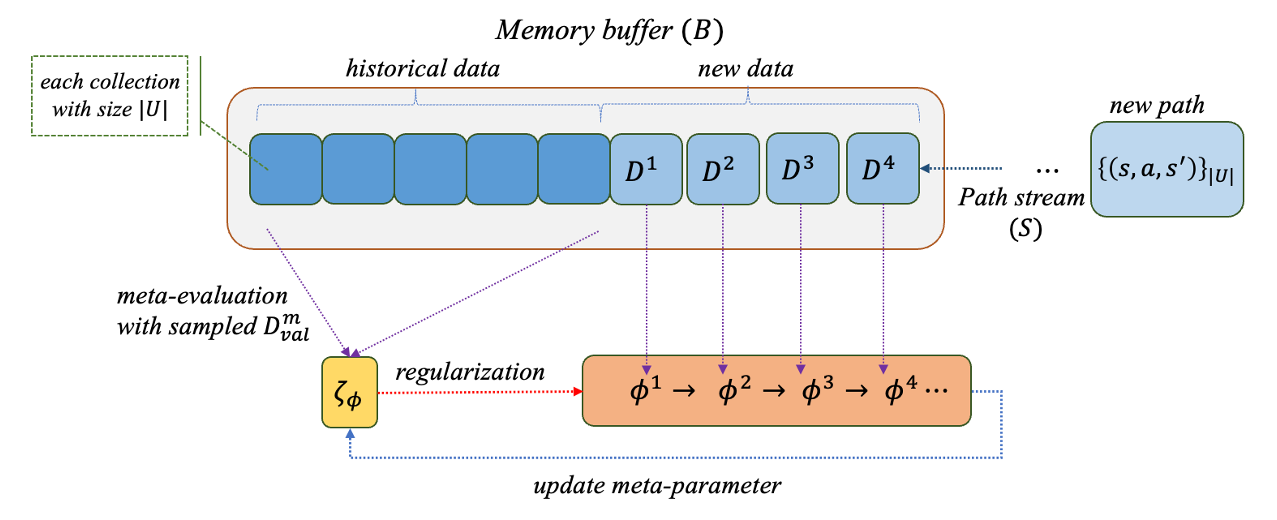}}
    {Diagram of the FOML Framework for Dynamic Embedding Updates  \label{fig:dynamically update}}
    {The fully online meta-learning (FOML) framework is employed to update the parameters of the encoder at the start of each validation window (see Figure \ref{fig:backtesting}). Each update incorporates new data (a block in the memory buffer) while also leveraging previous knowledge. FOML leverages regularization to facilitate the quick adaptation of the parameter $(\phi,\theta)$ to the new task.
    %The FOML framework is used to update the parameters of the encoder at the beginning of each validation window (see Figure \ref{fig:backtesting}). Each update uses new data (a block in the memory buffer) and references previous knowledge. FOML uses regularization to speed up the adaptation of the parameter $(\phi,\theta)$ to the new task, as long as past tasks are reasonably representative of prior tasks. More details can be found in \cite{rajasegaran2022fully}. 
    }
\end{figure}

The idea behind meta-learning is to train a base model that can quickly adapt to different scenarios, allowing updates with very few samples when faced with new situations.
In our framework, we first train a base generative autoencoder $\Gamma_{\zeta_\phi}$ and $G_{\zeta_\theta}$ using historical data by minimizing the loss $\mathcal{L}(\zeta_\phi,\zeta_\theta)$ as defined in Equation \eqref{eq: generative loss}. We then treat every $|U|$ periods as a new scenario and use the latest observed data within these $|U|$ periods to update the autoencoder. We illustrate the framework in Figure \ref{fig:dynamically update}. 
%Considering the fact that $\Gamma_\phi$ and $G_\theta$ are trained together, for notation brevity, we concatenate their parameters $(\phi,\theta) = (\phi^\top,\theta^\top)^\top$.

We first collect a set of data $H=\{(\boldsymbol{s_{i}},\boldsymbol{a_{i}},\boldsymbol{s^\prime_{i}})\}_{i=1}^{|H|}$, and use it to  train a base autoencoder parameterized by $(\phi,\theta) = \zeta := (\zeta_\phi,\zeta_\theta)$. The data for training the base autoencoder can be real trading logs or simulated trading paths on historical data. During the online update phase, we update the autoencoder every $|U|$ periods with the latest data. The new data $\mathcal{D}^j=\{(\boldsymbol{s}_{i}^j,\boldsymbol{a}_{i}^j,\boldsymbol{s}^{\prime j}_{i})\}_{i=1}^{|U|}$of size $|U|$ is continuously added to a memory buffer, where the superscript $j$ indicates the $j^{\text{th}}$ stream. The online update step relies on the most recent information in the buffer, which contains the latest market knowledge. 
To update the encoder, we use the latest $\mathcal{D}^{j}$. This data stream is then split into a training set $\mathcal{D}^{j}_{tr}$ and a validation set $\mathcal{D}^{j}_{val}$. 

% During the exploration, the stream of visited data paths $S=\{(\boldsymbol{s_{t}},\boldsymbol{a_{t}},\boldsymbol{s^\prime_{t}})\}$ of size $|S|$ is continuously added to the memory buffer. When updating the encoder, the online update step is based only on the latest information in the buffer, which contains more recent knowledge about the current market. For each update, the new data stream contains $|S|$ consecutive path pairs (which we call tasks in meta-learning) and represents the $i^\text{th}$ sequence as $\mathcal{D}^{i}$. We then divide $\mathcal{D}^{i}$ into a training set $\mathcal{D}^{i}_{tr}$ and a validation set $\mathcal{D}^{i}_{tr}$. 

The process of updating the embedding involves transferring knowledge from the base parameter vector $\zeta$ to the online parameter vector $(\phi^j,\theta^j)$, which represents the $j$-th update. Specifically, online meta-learning uses prior knowledge $\zeta$ as a regularizer for the online parameter $(\phi,\theta)$. As suggested by  \cite{rajasegaran2022fully}, a squared error of the form $\mathcal{R}(\phi,\theta, \zeta)=\|(\phi^\top,\theta^\top)^\top-\zeta\|^{2}$ is chosen as the regularization term, securing that the new parameter $(\phi,\theta)^j$ do not change drastically. This results in the following online update for the encoder $\Gamma_\phi$ at each step $j$:
\begin{equation}\label{eq:FOML updating 1}
\begin{aligned}
\phi^{j}=&\phi^{j-1}-\alpha_{1} \nabla_{\phi^{j-1}}\left\{\mathcal{L}\left(\phi^{j-1},\theta^{j-1} ; \mathcal{D}_{t r}^{j}\right)+\beta_{1} \mathcal{R}\left(\phi^{j-1},\theta^{j-1}, \zeta\right)\right\}\\
&=\phi^{j-1}-\underbrace{\alpha_{1} \nabla_{\phi^{j-1}} \mathcal{L}\left(\phi^{j-1},\theta^{j-1} ; \mathcal{D}_{t r}^{j}\right)}_{\text {new-data direction update }}+\underbrace{2 \alpha_{1} \beta_{1}\left(\zeta_\phi-\phi^{j-1}\right)}_{\text{meta direction update}},
\end{aligned}
\end{equation}
where $\alpha_1$ and $\beta_1$ are the learning rates, and $\mathcal{L}\left(\phi^{j-1},\theta^{j-1} ; \mathcal{D}_{t r}^{j}\right)$ is the loss from \eqref{eq: generative loss} with all data $(\boldsymbol{s},\boldsymbol{a},\boldsymbol{s}')$ from $\mathcal{D}_{t r}^{j}$:
\begin{equation} \label{eq: meta loss}
    \mathcal{L}(\phi^{j-1},\theta^{j-1};\mathcal{D}_{t r}^{j}) = \sum_{i=1}^{|U|} \mathcal{C}\left(\boldsymbol{s}'_i,\mathbb{E}_{G_\theta(\hat{s}'_i|z_{s_i}\sim\Gamma_\phi(z_{s_i}|s_i),a_i)}\left[\boldsymbol{\hat{s}}'_i\right]\right).
\end{equation}

The new-path direction update fine-tunes the autoencoder to minimize the reconstruction loss in \eqref{eq: generative loss} for the newly visited path. This step adapts the embeddings to current market dynamics, capturing new market patterns. The meta direction update acts as a penalty term to prevent the new parameters from changing drastically, ensuring a stable learning process for downstream RL tasks.  The decoder $G_\theta$ is updated using a similar logic as in \eqref{eq:FOML updating 1} by replacing $\phi$ with $\theta$.

We also incorporate the new information into the prior knowledge, by updating $\zeta$ using the following equation:
\begin{equation}\label{eq:FOML updating 2}
\zeta=\zeta-\alpha_{2} \nabla_{\zeta}\mathcal{L}\left(\phi^j,\theta^j ; \mathcal{D}_{val}^{m}\right)-2\alpha_{2}\beta_{2} \sum_{k=0}^{J} \left(\zeta- (\phi^{j-k},\theta^{j-k})\right),
\end{equation}
where $\mathcal{D}_{val}^{m}$ is a set of randomly selected data from the memory buffer $\mathcal{D}_\text{buffer}$, and $J$ indicates that the update considers its previous $J$ updates.

Once the autoencoder has been updated with the new parameters $(\phi^j,\theta^j)$, the RL agent in the $j+1$-th step makes a trading action based on the new embedding from $\Gamma_{\theta^j}$. Importantly, the portfolio allocation policy follows the form $\pi(\boldsymbol{a}|\boldsymbol{z}_s)$ and therefore the dynamic update of the encoder results in an updated embedding state $\boldsymbol{z}_s$ for the same raw state $\boldsymbol{s}$, which leads to different trading actions under the updated embedding.

\subsection{Dynamic Embedding Reinforcement Learning (DERL)} 
\label{sec: main of Embed then RL}

This section introduces the end-to-end Dynamic Embedding Reinforcement Learning (DERL) framework, which integrates dynamic embedding with a reinforcement learning algorithm. We also provide an implementation using the Wasserstein autoencoder as the generative encoder and the TD3 algorithm as the reinforcement learning component.

\begin{figure}[htbp]
    \FIGURE
    {\includegraphics[scale=0.85]{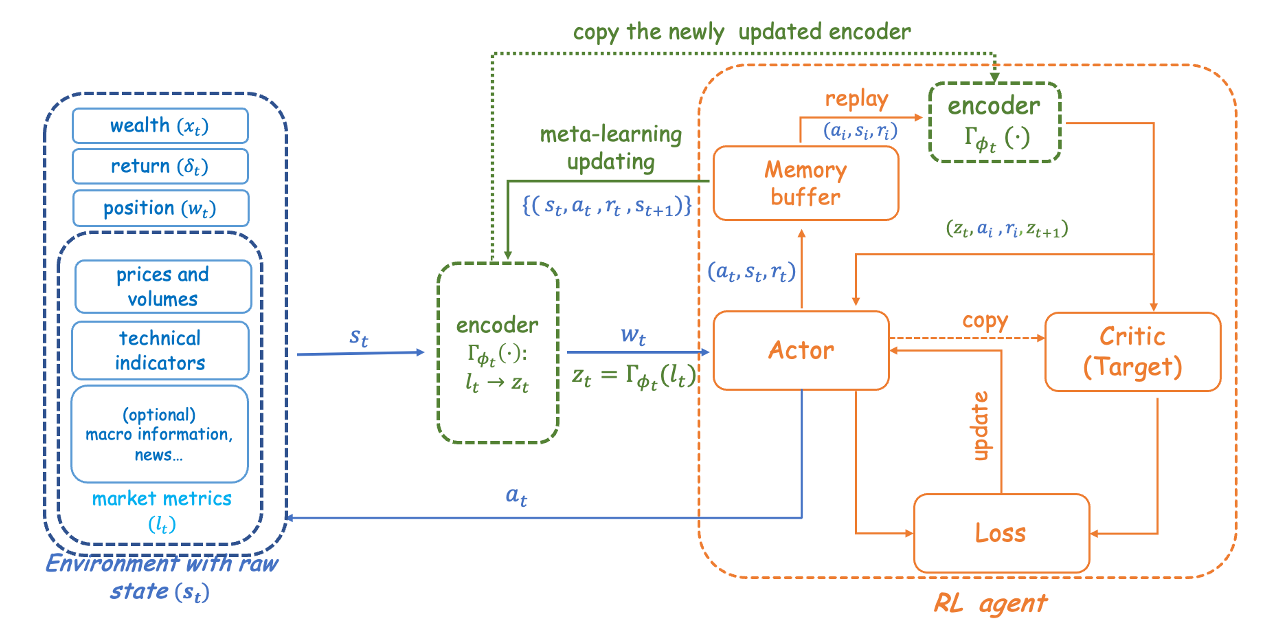}}
    {The DERL Framework. \label{fig:pipeline}}
    {The state $\boldsymbol{s_t}$ is first encoded into a low-dimensional latent state $\Gamma_{\phi_t} (\boldsymbol{s_t})$. The agent then learns the policy based on this encoded state. The experienced paths $(\boldsymbol{s},\boldsymbol{a},r,\boldsymbol{s^\prime})$ are saved in a memory buffer. Each time the RL agent is trained, $n$ paths $\left\{(\boldsymbol{s},\boldsymbol{a},r,\boldsymbol{s^\prime})\right\}_n$ are randomly sampled from the memory buffer and encoded into $\left\{(\Gamma_{\phi_t}(\boldsymbol{s}),\boldsymbol{a},r,\Gamma_{\phi_t}(\boldsymbol{s^\prime})\right\}_n$ to update the RL parameters. To capture the latest transition dynamics, the embedding is periodically updated with data from the memory buffer using online meta-learning with the latest path. In the pipeline, blue parts indicate the flow of raw states and actions, green parts indicate the embeddings, and orange parts represent computations and updates within the reinforcement learning agent.}
\end{figure}

The DERL framework uses a generative autoencoder to encode the current state into a low-dimensional latent state, which is then used to train the RL agent. The framework is designed to be continuously updated using online meta-learning to adapt to changing market conditions. 
Figure \ref{fig:pipeline} illustrates our framework, and the detailed algorithm implementation is provided in Algorithm \ref{alg: TD3}.

To train the RL agent in DERL, we save all observed tuple $(\boldsymbol{s},\boldsymbol{a},r,\boldsymbol{s^\prime})$ in a memory buffer $\mathcal{D}_\text{buff}$. The state $\boldsymbol{s}$ is first encoded into a lower-dimensional latent state $\boldsymbol{z_s}$ using the encoder $\Gamma_{\phi}$ trained in the generative autoencoder (as discussed in \S \ref{sec:generative models and autoencoders}). The RL agent learns the policy based on this encoded state $\boldsymbol{z_s}$. During each training iteration, we randomly sample $n$ data points $\left\{(\boldsymbol{s},\boldsymbol{a},r,\boldsymbol{s^\prime})\right\}_n$ from the memory buffer and encode them into $\left\{(\boldsymbol{z_s},\boldsymbol{a},r,\boldsymbol{z_{s^\prime}})\right\}_n$ with current encoder $\Gamma_\phi$ for training the RL agent. Additionally, we continuously update the encoder with new data from the memory buffer using online meta-learning to capture the latest transition dynamics (as discussed in \S \ref{sec:fine-tuning}).

% We note that DERL is a general framework for dynamic portfolio allocation and can work with various generative autoencoders and reinforcement learning algorithms. In this study, we use variational autoencoders and Wasserstein autoencoders as the generative models, and the Twin Delayed Deep Deterministic Policy Gradient (TD3) algorithm as the reinforcement learning method, which we briefly discuss subsequently.

\begin{comment}
In the following sections, we present an example of the realized framework using variational autoencoder or Wasserstein autoencoder as the generative autoencoder in \S \ref{sec:vae and wae}, and  the twin delayed deep deterministic policy gradient (TD3) algorithm as the reinforcement learning algorithm in \S \ref{sec:TD3}. 
    
\end{comment}

\subsubsection{Embedding with WAE} \label{sec:vae and wae}

This section briefly overviews how Wasserstein autoencoder (WAE) \citep{tolstikhin2018wasserstein} is used as a generative model in the DERL framework. WAE minimizes the Wasserstein distance between the encoded distribution and a known prior distribution, mapping the data distribution to the prior.

To train the generative autoencoder, we minimize the loss defined in \eqref{eq: generative loss}. For WAE, when encoding a state, we sample the embedded variable $\boldsymbol{z_{s}}\sim \Gamma_\phi(\boldsymbol{s})$, and when reconstructing in WAE, we use a deterministic decoder, which means $G_\theta = \delta(\cdot\mid\boldsymbol{z_{s}}, \boldsymbol{a})$ and can be simplified as $G_\theta(\boldsymbol{z_{s}}, \boldsymbol{a})$. Then, the loss of training defined in \eqref{eq: generative loss} can be approximated through the empirical loss:
\begin{equation}\label{eq:WAE-MMD}
\mathcal{L}_{\text{WAE-MMD}}\left(\phi, \theta\right)=\underset{\left(\boldsymbol{s}, \boldsymbol{a}, \boldsymbol{s^{\prime}}\right)\in\mathcal{D}\text{buffer} }{\sum}\left[ \mathcal{C} (\boldsymbol{s}^\prime, G_\theta(\boldsymbol{z_{s}}, \boldsymbol{a}))
+\lambda \mathcal{L}_{\mathrm{MMD}}\left(\Gamma_{\phi}\left(\boldsymbol{s}\right) | \mathcal{P}_{\text{prior}}\right)\right],
\end{equation}
where the MMD behind WAE indicates that the maximum mean discrepancy (MMD) loss is used to measure the distribution distance. The loss function consists of two components: reconstruction loss and discrepancy loss. The reconstruction loss $\mathcal{C}(\cdot,\cdot)$ measures the difference between the original input and the reconstructed output, while the discrepancy loss $\mathcal{D}(\cdot,\cdot)$ measures the difference between the learned latent space and a pre-defined prior distribution $\mathcal{P}_\text{prior}$. For practice, the prior distribution is usually set as standard multivariate Gaussian distribution: $\mathcal{P}_\text{prior} = \mathcal{N}(\boldsymbol{0}, \boldsymbol{I}).$ The discrepancy loss $\mathcal{D}_\text{MMD}$ is defined using a positive-definite reproducing kernel $k: \mathcal{Z} \times \mathcal{Z} \rightarrow \mathbb{R}$, and computed as:
\begin{equation}
\mathcal{L}_{\text{MMD},k}\left(\Gamma_{\phi,\boldsymbol{s}}, \mathcal{P}_\text{prior}\right)=\left\|\int_{\mathcal{Z}} k(\boldsymbol{z}, \cdot) d \Gamma_{\phi,\boldsymbol{s}}(\boldsymbol{z})-\int_{\mathcal{Z}} k(\boldsymbol{z}, \cdot) d \mathcal{P}_\text{prior}(\boldsymbol{z})\right\|_{\mathcal{H}_{k}},
\end{equation}
where $\mathcal{H}_{k}$ is the reproducing kernel Hilbert space (RKHS) of the real-valued function that maps $\mathcal{Z}$ to $\mathbb{R}$, and $\Gamma_{\phi,\boldsymbol{s}}$ indicates the learned latent distribution of $\boldsymbol{z}$ for given raw state $\boldsymbol{s}$.

For the implementation algorithm for training a WAE as a market state encoder, see Algorithm  \ref{alg:WAE} in the E-Companion for more details.

\subsubsection{TD3 Reinforcement Learning Algorithm} \label{sec:TD3}  

\begin{comment}
    We use the Twin Delayed DDPG (TD3) algorithm, which includes a memory buffer to improve sample efficiency by reusing previously encountered data. The RL parameters are updated by randomly sampling $n$ paths from the memory buffer, encoding them into $\left{(\Gamma_\phi(\boldsymbol{s}),\boldsymbol{a},r,\Gamma_\phi(\boldsymbol{s^\prime}))\right}_n$. The encoder is updated periodically using data from the memory buffer.

    After training an embedding to map high-dimensional original state to a low-dimensional latent space, we train the RL agent to learn how to trade using the embedded states $\boldsymbol{z_s}$. Since the encoder is fixed during the RL training and validation steps, the embedding allows us to represent the MDP and value function, discussed in \S \ref{sec: Main framework}, using the low-dimensional latent states $\boldsymbol{z_s}$ instead of the original high-dimensional states $\boldsymbol{s}$. 
\end{comment}

The Twin Delayed Deep Deterministic Policy Gradient (TD3) algorithm builds upon the Deep Deterministic Policy Gradient (DDPG) algorithm \citep{lillicrap2015continuous}. DDPG is a model-free off-policy algorithm that uses deep neural networks to learn policies in continuous action spaces. TD3 addresses issues such as overestimation bias and learning instability by incorporating three key improvements: \textit{double Q-learning}, \textit{delayed policy updates}, and \textit{target policy smoothing}.

Double Q-learning mitigates overestimation bias by using two critic networks to estimate the value of the next state and taking the minimum value between them. Delayed policy updates improve learning stability by updating the policy network less frequently than the value networks. Target policy smoothing adds noise to the target action to make the value estimation more robust to slight changes in action selection, reducing the variance in value estimates.

% \ins{In Figure \ref{fig:pipeline}, we use actor-critic model-free algorithms as the RL agent, which is the architecture of many state-of-the-art RL algorithms \citep{fujimoto2018td3,qu2022scalable,zhang2022actor}. The \textit{actor} generate buy-sell actions based on the policy $\pi$, while the \textit{critic} evaluates the corresponding value function. Without losing generality, this architecture can also be simply modified to other RL algorithms like the famous Q-learning.}

The TD3 algorithm uses six neural networks to approximate the value function and generate policies. These include two critic networks, $Q_{\nu_{1}}$ and $Q_{\nu_{2}}: \mathcal{Z}\times\mathcal{A}\rightarrow\mathbb{R}$, parameterized by $\nu_{1}$ and $\nu_{2}$, respectively, which evaluate the (state-action) value function. The actor network, $\pi_\iota: \mathcal{Z}\rightarrow \mathcal{A}(\boldsymbol{s})$, generates an allocation action for a given state. 
Additionally, there are corresponding target networks, $Q_{\nu_{1}^\prime}, Q_{\nu_{2}^\prime}$ for the critic networks, and $\pi_{\iota^\prime}$ for the actor network. 
The target networks are delayed copies of their original networks, providing more stable and reliable targets for the critic networks.

At each step, the TD3 agent interacts with the market according to the policy from the actor network $\pi_\iota$ and stores its experiences in a replay buffer. The algorithm then uses a batch of experiences to update the critic networks, predicting the value of taking an action in a given state. The actor network is updated using the predicted values from the critic networks to determine the best allocation action to take in a given state.
The target networks are updated slowly by copying the weights from the online networks at a small update rate, ensuring that the learning process remains stable. This updating process continues until the agent's performance converges to an optimal level, indicating that the agent is making profitable investment decisions.

The two critic networks are updated simultaneously by minimizing the mean squared error between a target value and the estimated state-action value: 
\begin{equation}
    \nu_{i} = \operatorname{argmin}_{\nu_{i}} N^{-1} \sum\left(y-Q_{\nu_{i}}(\boldsymbol{z_s}, \boldsymbol{a})\right)^{2} \quad (i=1,2),
\end{equation}
where $y = r+\gamma \min _{i=1,2} Q_{\nu_{i}^{\prime}}\left(\boldsymbol{z_s}(\boldsymbol{s^{\prime}}), \tilde{\boldsymbol{a}}\right)$ represents the minimum value of the two target networks' outputs. In financial portfolio management, TD3 uses its actor networks to take investment actions in a given embedding market state. The target value is calculated using a pair of target Q-value networks that predict the expected utility.
% TD3 employs a target policy smoothing technique to improve the stability and convergence of the policy gradient methods, which allows it to explore the continuous action space more effectively, improve learning speed, and converge to better investment policies. This technique helps the algorithm to make more informed investment decisions and maximize returns for the portfolio.

TD3 updates the actor network using the policy gradient method. The update rule involves computing the gradient of the expected state-action value with respect to the actor network parameters. This gradient measures how changes in the actor network parameters affect the expected utility. The actor network is then updated by taking a step in the direction that increases the expected utility, enhancing its ability to select actions that maximize the portfolio's Sharpe ratio. This process continues until the algorithm converges.
\begin{equation}
   \iota=\iota-\alpha_\iota \nabla_{\iota} J(\iota)=\iota-\alpha_\iota \left.N^{-1} \sum \nabla_{\boldsymbol{a}} Q_{\nu_{1}}(\boldsymbol{z_s}, \boldsymbol{a})\right|_{\boldsymbol{a}=\pi_{\iota}(\boldsymbol{z_s})} \nabla_{\iota} \pi_{\iota}(\boldsymbol{z_s}).
\end{equation}

Finally, the algorithm updates the target networks with low frequency, by softly interpolating their parameters with those of the online networks 
\begin{equation}
\begin{aligned}
\nu_{i}^{\prime} \leftarrow \tau \nu_{i}+(1-\tau) \nu_{i}^{\prime},
\end{aligned}
\end{equation}
where the $\tau$ is a soft-update coefficient that controls the speed of the update. This approach ensures that the learning process remains stable and the policy gradually converges.

For more details on the TD3 algorithm and its implementation, please refer to \S \ref{ec:td3} and Algorithm \ref{alg: TD3}.

\section{An Empirical Study of U.S. Equities}\label{sec:numerical}
In this section, we assess the out-of-sample performance of the DERL framework using thirty years of U.S. equities data, comparing it with alternative models. We outline the data and evaluation design in \S \ref{sec: Experiment detail}, detail the implementation parameters in \S \ref{sec: parametric}, and demonstrate the framework's performance against baseline models in \S \ref{sec:numerical}. We analyze portfolio performance using factor analysis and lasso regression to decode the return components and decision-making patterns of the DERL agent in \S \ref{sec:interpretibility}. Finally, ablation studies exploring the impact of embedding and dynamic updating are discussed in \S \ref{sec:ablation}.

%In \S \ref{sec: Experiment detail} and \ref{sec:rolling window}, we describe the data and the rolling-window portfolio performance evaluation scheme used in our empirical study. In \S \ref{sec: numerical results}, we present and analyze the out-of-sample portfolio performance. 
% Our framework is implemented in Python 3.8 and executed on two Nvidia GeForce RTX 3090 GPUs with 24GB of RAM.

\subsection{Data and Empirical Design} \label{sec: Experiment detail}
We evaluate our model performance using the top 500 stocks by market value, which are actively traded. The trading information for each constituent stock, including daily open (O), high (H), low (L), close (C) prices, trading volumes (V), and returns, is collected from the CRSP (Center for Research in Security Prices) database, covering the period from January 1, 1990, to December 31, 2022. We incorporate various technical indicators for each constituent stock, including Simple Moving Averages (SMA-21-day/42-day/63-day), Exponential Moving Averages (EMA-21-day/42-day/63-day), Moving Average Convergence Divergence (MACD), Relative Strength Index (RSI-21-day/42-day/63-day), Bollinger Bands (BOLL), Commodity Channel Index (CCI-21-day/42-day/63-day), Average Directional Index (ADX-21-day/42-day/63-day), On-Balance Volume (OBV), Stochastic Oscillator, Chaikin Money Flow (CMF), Accumulation/Distribution Line (ADL), and Williams \%R. Additionally, we include two market-level variables: the daily U.S. Treasury spot rate and the USD/EUR exchange rate. Thus, for the experiment with the top 500 stocks, the raw state dimension is $15,506$, and the action dimension is a vector of size $D=501$.

%Similarly, for the CSI 500, the state dimension is $15 \times 359 + 3 = 5,388$ and the action dimension is $359$.

% We collect daily open (O), high (H), low (L), close (C) prices, and trading volumes (V) for each constituent stock and use the close price as the trading price $\boldsymbol{p}$, while the other three prices are left in the auxiliary information $\boldsymbol{l}$. Other auxiliary information includes technical indicators for each constituent stock: simple moving average (SMA-21-day and SMA-63-day),  moving average convergence divergence (MACD), relative strength index (RSI-21-day and RSI-63-day), Bollinger bands (BOLL), commodity channel index (CCI-21-day), average directional index (ADX-21-days), and the daily US treasury spot rate and the exchange rate of USD/CNY. Thus, for the experiment of trading S\&P 500 constituents, the state dimension is 
% $15 \times 471 + 3 =7069$, and the action is a vector of size $471\times1$. For the CSI 500, the state dimension is $15 \times 359 + 3 = 5388$ and the action dimension is $359$.

%\subsubsection*{Rolling-window Backtesting Details.}\label{sec:rolling window}
%\subsection{Methods for model evaluation} \label{sec:rolling window}
\subsubsection{Segments of Back-testing Period}
We conduct a thirty-year backtest of our framework using data from January 1, 1993, to December 31, 2022. Due to fluctuations in the market value of equities over time, we segment the backtesting timeline into six disjoint five-year periods: 1993-1997, 1998-2002, 2003-2007, 2008-2012, 2013-2017, and 2018-2022. At the beginning of each period, we establish a new portfolio consisting of the top 500 market-value stocks. For each period, data from the previous three years served as the training set for our model, with the following five years dedicated to applying and iteratively updating our trading strategy on a rolling basis.

For instance, our analysis for the first period from 1993 to 1997 is based on the top 500 stocks as of the last trading day of 1992. The data for these stocks from the beginning of 1990 through the end of 1992 are used for model training, with trading activities commencing at the beginning of 1993. At the start of 1998, we construct a new portfolio for the subsequent period based on the top 500 equities as of the end of 1997. The data from 1995 to 1997 serve as the training phase, with the new trading strategy launching at the start of 1998. We present the testing periods, the corresponding training windows, and the portfolio components in Table \ref{tab:backtesting_segments}.

\subsubsection{Rolling-window Backtesting in Segment}
\begin{figure}[h]
    \FIGURE
    {\includegraphics[scale=0.75]{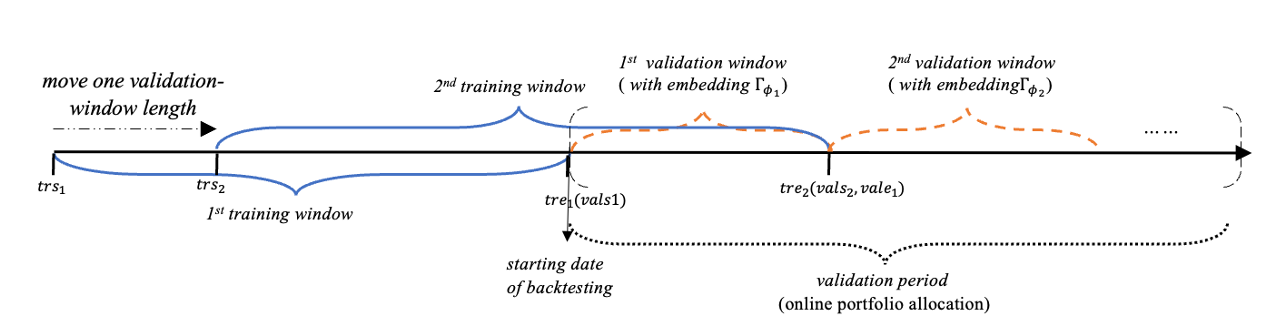}}
    {Rolling-window backtesting.  \label{fig:backtesting}}
    {We use a rolling window-based backtesting process inside each backtesting segment to evaluate the performance of the trading strategy. In the figure, $\text{trs}_i$ represents the start date of the $i^{\text{th}}$ training window, and $\text{tre}_i$ represents its end date. Similarly, $\text{vals}_i$ and $\text{vale}_i$ denote the start and end dates of the validation window, with $\text{vals}_i=\text{vale}_{i-1}=\text{tre}_i$
    %We deploy a rolling window-based backtesting process to check the performance of the trading strategy. In the figure, $\text{trs}_i$ denotes the date when the $i^{\text{th}}$ training starts, and $\text{tre}_i$ denotes its end date. Similarly, $\text{vals}_i$ and $\text{vale}_i$ denote the start and end dates of the validation window. We have that $\text{vals}_i=\text{vale}_{i-1}=\text{tre}_i$.
    }
\end{figure}

For each segment, we follow a fixed-length rolling window scheme shown in Figure \ref{fig:backtesting}, which is similar to the moving-window approach described in \cite{fama1988permanent}. 
%The online portfolio allocation starts from $vals_1$ in Figure \ref{fig:backtesting}. 
We divide each segment period into non-overlapping, consecutive validation windows (such as the $1^{\text{st}}$ and $2^{\text{nd}}$ validation windows in Figure \ref{fig:backtesting}). The length of these windows is determined by how frequently we update our embedding and RL parameters. 
Our approach to updating the encoder $\Gamma_{\phi^j}$ for the $j^{\text{th}}$ validation window follows the online meta-learning framework introduced in \S \ref{sec:fine-tuning}, and the approach to updating the RL agents follows the method introduced in \S \ref{sec:TD3} (detailed in Algorithm \ref{alg: TD3}).% In addition, a stop-loss mechanism is implemented for stability during trading execution, similar to \cite{yang2020deep}, where all positions are closed for b when the daily market turbulence exceeds the 95 percentile of historical turbulence. 
% \textcolor{blue}{We use the past $d_{\text{train}}$ days of data, represented in Figure \ref{fig:backtesting} as the $j^{\text{th}}$ \textit{training window}, for the $j^{\text{th}}$ validation window to update the RL agent for the $j^{\text{th}}$ validation window. }

In each training window $j$, we use the parameters inherited from the previous validation window $j-1$ as a starting point for our RL agent, which contains previously learned knowledge. The agent then explores and learns for various iterations from the training start date, $\text{trs}_j$, to the training end date, $\text{tre}_j$. The data tuples $(\boldsymbol{s},\boldsymbol{a},\boldsymbol{s^\prime})$ visited during this period are saved for updating the embedding encoder $\Gamma_{\phi^j}$ using the online meta-learning updating equations (\ref{eq:FOML updating 1}) and (\ref{eq:FOML updating 2}) via gradient information. After updating the encoder $\Gamma_{\phi^j}$ and the RL agent, we conduct backtesting in the $j^{\text{th}}$ validation window.

In the backtest, the training window starts on the first trading day of each segment, respectively ($\text{trs}_1$=January 1, 1990, 1995, 2000, 2005, 2010, and 2015) and the first validation window starts on the first day of 1993, 2003, 2013, respectively ($\text{vals}_1$= January 1, 1993, 1998, 2003, 2008, 2013 and 2018). The length of each validation window is 42 days, and the validation period for each segment ends on the last day of 1997, 2002, 2007, 2012, 2017, and 2022,
 respectively. The entire 30-year backtesting horizon, considering 252 trading days every year, we have in total 180 validation windows. Besides, a transaction cost rate of 0.1\% is applied to the total value of each trade. 

% At the beginning of each backtesting period, the agent is provided with an initial cash balance of $1\times10^7$ to account for capital constraints. To simplify the trading strategy, all trades are based on the close price, as is common in algorithmic trading literature. 

\subsection{Experimental Parameters and Configuration} \label{sec: parametric}
This section presents the parameters of all three components (WAE, FOML, and TD3) in the DERL framework introduced in \S\ref{sec: main of Embed then RL}, and briefly discusses the computation time and complexity of each algorithm.
\subsubsection*{WAE Parameters and Configurations}
For the training of the embedding layer, we initially train the encoder following Algorithm \ref{alg:WAE}. The batch size $n$ is set to 40, and the prior distribution $P_Z$ for WAE is assumed to be standard Gaussian. The layer sizes of the encoder are $\dim (S), 512, 512, \dim (Z)$, and the auxiliary decoder is a multi-layer perceptron (MLP) with layer sizes $\dim (Z) + \dim (A), 512, 512, \dim (S)$. We use the inverse multiquadratics kernel $k(x,y) = \frac{ d_z^2}{d_z^2+\|x-y\|^2_2 }$, and the regularization parameter $\lambda$ is set to 2. 
In the experiments the embedding size $\dim(\mathcal{Z})$ is set to 500. We have also tested other embedding sizes from 50 to 2000 and found that 500 is an effective choice. Embedding sizes between 300 and 600 provided similar results.

\subsubsection*{FOML Parameters and Configurations}
When updating the embedding according to the FOML algorithm introduced in \S \ref{sec:fine-tuning}, the learning rates are set to $\alpha_1=0.0001, \beta_1=0.001, \alpha_2=0.0005, \beta_2=0.005$. In the first training period of the meta-learning model, we perform 6 million iterations of backpropagation for the loss and follow the updates outlined in \S \ref{sec:fine-tuning}. The update frequency is set to $|U|=42$, which is also the validation window length shown in Figure \ref{fig:backtesting}. This indicates that we dynamically update the embedding every 42 days.

\subsubsection*{TD3 Parameters and Configurations}
For the RL network used in the backtest, the state-action value function $Q(\boldsymbol{z_s},\boldsymbol{a})$ for the TD3 agent is implemented as a three-hidden-layer fully connected neural network (FCN) with a rectified linear unit (ReLU) activation function. The layer sizes are $\dim (Z)+\dim (A), 256, 256, 256, 1$, from the input layer to the output layer, as suggested by \cite{fujimoto2018td3}. The actor's policy network $\pi_\iota$ is also an FCN with layer sizes $\dim (Z), 256, 256, 256, \dim (A)$. The discount factor $\gamma$ is set to $0.999$, the learning rate of the policy network $\alpha_\iota = 0.0002$, the soft-update parameter $\tau= 0.005$, and the target network is updated every five trading days. 

\subsubsection*{Experiment Implementation and Time Complexity}
We implement the DERL framework with Python 3.8 and PyTorch on four NVIDIA GeForce RTX 3090 GPUs. All parameters of neural networks are initialized with normal initialization with a standard deviation of 0.001. 

To illustrate the time complexity of training each key component in our experiment, we conduct the experiments 20 times and calculated the average time for each part. The initial training stage of embedding with 6 million randomly collected paths takes approximately 15.3 hours. Each dynamic update of the embedding takes about 5.15 minutes. Training the RL agent within one training window (42-day) takes 11.2 minutes. To fully execute one 30-year back-testing path, including training the RL agent, updating embeddings with meta-learning, and executing portfolio allocation actions based on the learned policy, the total estimated time is approximately 45.7 hours\footnote{This estimate assumes the use of parallel training techniques on GPUs, which may save some time by conducting all parts sequentially.}.

\subsection{Out-of-sample Performance} \label{sec: numerical results}

Table \ref{tab: SP500 summarize performances} presents the out-of-sample investment performance of our DERL agent, an RL model utilizing dynamic embedding within our framework. For comparison, we also detail the performance of the two-step PTO method using an MLP model, following the methodology outlined in \cite{gu2020empirical}, and two standard benchmarks—the value-weighted and equal-weighted portfolios. Key metrics reported include annualized mean, standard deviation (STD), skewness (Skew), kurtosis (Kurt), Sharpe ratio (SR), and Sortino ratio (ST) for each portfolio's returns.
%To test the null hypothesis that the  value- or equal-weighted portfolios can have significant better investment performance than the DERL model, we follow \cite{DPUV2013} to use the bootstrap method. Specifically, we generate $M$ pairsof daily portfolio returns, where $M$ is the number of the out-of-sample observations, by drawing randomly with replacement from the realized portfolio returns of the DERL portfolio and an alternative portfolio (value- or equal-weighted portfolio). Suppose the difference of the performance measures between the two strategies is $D_{i}$ for the $i$-th trial. By repeating the above procedure for $B=10,000$ times, we can obtain a series of $D_{i},i=1,\cdots ,B$. We then can obtain the one-sided p-value for the null hypothesis that the DERL portfolio underperforms the alternative portfolio by calculating the frequency that $D_{i}$ is lower than zero. 

\begin{table}[htbp]
\setlength{\tabcolsep}{13pt}
\TABLE
{Out-of-sample performance\label{tab: SP500 summarize performances}}
{\begin{tabular}{cccccccc}
\toprule
           &            &       Mean &        STD &       Skew &       Kurt &         SR &         ST \\
\midrule
           \multicolumn{8}{l}{Panel A: Full samples (N=7550)}\\
\midrule

            DERL &            &    0.1481  &    0.1423  &    1.7526  &   36.7457  &    1.0407  &    1.6200  \\

        2Step &            &   0.1203 &     0.2254 &    -0.2683 &    20.6988 &     0.5338*** &     0.7508*** \\

        VW &            &    0.0895  &    0.1864  &   -0.1871  &   13.3305  &  0.4802*** &  0.6769*** \\

         EW  &            &        0.1369 &     0.1973 &    -0.2129 &    15.4837 &      0.6940*** &     0.9862*** \\

\midrule
       \multicolumn{8}{l}{Panel B1: Subsamples (1993-2002, N=2519)}\\
\midrule           

      DERL &            &    0.1451  &    0.1184  &    3.4429  &   69.3321  &    1.2257  &    2.0375  \\

        2Step &            &    0.0687 &     0.1801 &    -1.3966 &     21.781 &     0.3814*** &     0.5147*** \\ 

        VW &            &    0.0851  &    0.1746  &   -0.0328  &    6.6929  &  0.4873*** &  0.7012*** \\

         EW  &            &       0.1327 &     0.1526 &    -0.0833 &     7.7603 &     0.8697** &     1.2592***  \\

\midrule
\multicolumn{8}{l}{Panel B2: Subsamples (2003-2012, N=2517)}\\
\midrule           

       DERL &            &    0.1582  &    0.1719  &    1.3486  &   29.6616  &    0.9206  &    1.3958  \\

        2Step &            &     0.1555 &     0.2804 &     0.0739 &    16.9489 &     0.5547*** &     0.7932*** \\

        VW &            &    0.0695  &    0.2074  &   -0.0528  &   13.3492  &  0.3351*** &  0.4708*** \\

         EW  &            &        0.1367 &     0.2399 &    -0.0674 &    11.9524 &     0.5697*** &     0.8105*** \\

\midrule
\multicolumn{8}{l}{Panel B3: Subsamples (2013-2022, N=2514)}\\
\midrule
        DERL &            &    0.1410  &    0.1312  &    1.1379  &   19.2975  &    1.0743  &    1.6668  \\

        2Step &            &      0.1369 &     0.2036 &    -0.3585 &    19.0438 &     0.6724*** &     0.9487*** \\

        VW &            &    0.1139  &    0.1754  &   -0.5463  &   18.3967  &   0.6497** &  0.9027*** \\

        EW   &            &        0.1414 &     0.1898 &    -0.5407 &    20.5865 &     0.7452** &     1.0455*** \\

\midrule
\multicolumn{8}{l}{Panel C1: Low volatility regime (VIX<17.91, N=3371)}\\
\midrule
      DERL &            &    0.2369  &    0.0841  &    0.1265  &    4.6564  &    2.8171  &    4.6230  \\

        2Step &            &     0.3113 &     0.1123 &     0.0368 &     4.4234 &     2.7724 &     4.4968 \\

        VW &            &    0.2805  &    0.0961  &    0.0260  &    4.0436  &    2.9174  &    4.8036  \\

         EW  &            &    0.3077 &     0.1021 &    -0.0355 &     3.8647 &     3.0153 &     4.9277  \\

\midrule
\multicolumn{8}{l}{Panel C2: High volatility regime (VIX$\geq$17.91, N=3375)}\\
\midrule
      DERL &            &    0.0589  &    0.1827  &    1.7177  &   27.0572  &    0.3223  &    0.4960  \\

        2Step &            &   -0.0715 &     0.2979 &    -0.1297 &    13.4543 &    -0.2399*** &    -0.3312***\\

        VW &            &   -0.1019  &    0.2449  &   -0.0411  &    8.8306  & -0.4160*** & -0.5734*** \\

         EW  &            &      -0.0344 &     0.2593 &    -0.0794 &     10.269 &    -0.1328*** &     -0.1850*** \\

\bottomrule           
\end{tabular}
}
{\textit{Notes}: This table reports the out-of-sample performances of the DERL framework, two-step model, the value- and equal-weighted portfolios.
We report the annualized mean, standard deviation, skewness, kurtosis, Sharpe ratio, and Sortino ratio of the realized portfolio returns. We test the null hypothesis that the DERL framework produces a lower Sharpe or Sortino ratio than the alternative portfolio using the bootstrapping method \citep{DPUV2013}. Panel A presents the results for the full sample, Panels B1-B3 tabulate the results during three non-overlapping subperiods, and Panels C1-C2 present the results during low and high volatility regimes respectively. 
The asterisks *, **, and *** denote respectively the 10\%, 5\%, and 1\% level of statistical significance for the null that the full baseline model underperforms the alternative model. }
\end{table}

Panel A presents the results for the full sample. Compared with the other three models, the DERL agent achieves higher average returns, lower standard deviations, and significantly higher Sharpe and Sortino ratios. Additionally, the skewness of the DERL portfolio returns is positive, and the kurtosis is relatively small, indicating that the DERL framework effectively manages downside tail risks.

Panels B1-B3 further detail the performance of the four portfolios during different subperiods. Generally, the DERL agent consistently generates superior performance compared to the other three portfolios across different periods, with higher mean returns and lower standard deviations. Consequently, the DERL agent's portfolio outperforms the two-step, the value- and equal-weighted portfolios in terms of Sharpe and Sortino ratios. Overall, our empirical results in Table \ref{tab: SP500 summarize performances} suggest that, compared to the other three models, the DERL framework is more effective in optimizing investment returns while managing (tail) risk.

To shed light on the superior capability of the DERL framework in managing portfolio risk, Panels C1-C2 of Table \ref{tab: SP500 summarize performances} present the out-of-sample performance of DERL, two-step PTO (with MLP), value- and equal-weighted portfolios during different volatility regimes. We use the CBOE VIX, calculated based on the prices of S\&P 500 index options, to measure the market volatility. 
Panel C1 presents the results when the CBOE VIX value is lower than its historical median, or 17.91. It shows that the DERL agent enjoys returns with relatively lower mean and lower standard deviation. When using the Sharpe or Sortino ratio as the performance measure, the DERL agent does not significantly outperform the other two models.

As a comparison, when the VIX value is higher than its historical median, the DERL agent yields significantly higher Sharpe and Sortino ratios than the other two portfolios. Moreover, the DERL portfolio returns are right-skewed under both high and low volatility regimes. These results indicate that the DERL framework has superior capability in managing portfolio risk, especially during periods of market stress.

To determine whether the outperformance of the DERL framework can be explained by well-known factor models, we conduct a time-series analysis by regressing the out-of-sample DERL portfolio excess returns on the \cite{FAMA19933} three-factor model and the \cite{FAMA19933}-\cite{CAR1997} four-factor model\footnote{Regression results based on the \cite{fama2015} five-factor and \cite{hou2021} five-factor models show that these common factors cannot fully explain the DERL portfolio returns across different data samples, consistent with the main findings in Table \ref{tab:factor_analysis}. These additional results are available upon request.}. The estimation results are presented in Panels A and B, respectively.

\begin{table}[htbp]
\setlength{\tabcolsep}{7pt}
\TABLE{Factor analyisis of DERL portfolio\label{tab:factor_analysis}}
{
\begin{tabular}{lccccccccc}
\toprule
 & &  && \multicolumn{3}{c}{Subperiods} && \multicolumn{2}{c}{Volatility regimes} \\
  \cmidrule(r){5-7} \cmidrule(r){9-10}
& & Full sample && 1993-2002 & 2003-2012 &2013-2022 && Low & High \\
\midrule
\multicolumn{10}{l}{Panel A: Fama-French three-factor model}\\
 \midrule
  $\alpha$ &            &  0.0003*** &            &  0.0004*** &  0.0004*** &  0.0003*** &            &   0.0001** &  0.0005*** \\

           &            &   [6.3759] &            &   [3.8402] &   [4.1445] &   [2.8778] &            &   [2.1586] &   [4.7450] \\

    Market &            &  0.6380*** &            &  0.6204*** &  0.6485*** &  0.6086*** &            &  0.7499*** &  0.6198*** \\

           &            &  [50.5800] &            &  [36.0420] &  [34.6670] &  [22.8130] &            & [101.3600] &  [46.0740] \\

       SMB &            &  0.0859*** &            &   -0.0009  &  0.1327*** &  0.1293*** &            &  0.1098*** &   0.0625** \\

           &            &   [4.6880] &            &  [-0.0254] &   [3.8755] &   [4.9731] &            &  [10.3310] &   [2.5542] \\

       HML &            &  0.2698*** &            &  0.2507*** &  0.2905*** &  0.2409*** &            &  0.1915*** &  0.2873*** \\

           &            &  [13.8050] &            &   [8.3042] &   [5.6202] &  [10.9790] &            &  [15.4130] &  [12.5000] \\

   Adjusted $R^2$ &            &    0.7467  &            &    0.6327  &    0.7914  &    0.7712  &            &    0.7954  &    0.7424  \\

\midrule
\multicolumn{10}{l}{Panel B: Fama-French-Carhart four-factor model}\\
\midrule

  $\alpha$ &            &  0.0004*** &            &  0.0004*** &  0.0004*** &  0.0003*** &            &  0.0001*** &  0.0005*** \\

           &            &   [7.1540] &            &   [4.4872] &   [4.2597] &   [3.1853] &            &   [2.8061] &   [5.0347] \\

    Market &            &  0.6171*** &            &  0.6129*** &  0.6282*** &  0.5964*** &            &  0.7623*** &  0.5895*** \\

           &            &  [55.0700] &            &  [36.0830] &  [35.7780] &  [26.0730] &            &  [99.1520] &  [50.7890] \\

       SMB &            &  0.0840*** &            &    0.0092  &  0.1564*** &  0.0962*** &            &  0.1161*** &   0.0518** \\

           &            &   [4.8961] &            &   [0.2927] &   [4.7012] &   [3.7007] &            &  [10.1710] &   [2.2475] \\

       HML &            &  0.2177*** &            &  0.2409*** &  0.2131*** &  0.1876*** &            &  0.1671*** &  0.2221*** \\

           &            &  [12.4090] &            &   [7.8167] &   [4.9256] &   [8.5233] &            &  [13.1550] &  [10.7030] \\

       MOM &            & -0.1131*** &            & -0.1070*** & -0.1157*** & -0.1221*** &            & -0.1049*** & -0.1302*** \\

           &            &  [-9.1375] &            &  [-6.1052] &  [-5.2066] &  [-6.0202] &            &  [-7.8565] &  [-9.0832] \\

   Adjusted $R^2$ &            &    0.7593  &            &    0.6468  &    0.7988  &    0.7914  &            &    0.8065  &    0.7583  \\

\bottomrule
\end{tabular}
}
{\textit{Note}. Panels A and B of this table report time series regressions of the out-of-sample excess returns of the DERL portfolio on the 
Fama-French three-factor model, and the Fama-French-Carhart four-factor model respectively. The $t$-values with
Newey-West adjustments are reported in brackets, and the asterisks *, **, and *** denote the 10\%, 5\%, and 1\% level of statistical significance, respectively.}
\end{table}

Column 2 of Table \ref{tab:factor_analysis} shows that for the full sample, the DERL portfolio returns have significant loadings on the market factor, with the coefficient of the market factor exceeding 0.6 and being statistically significant at the 1\% level. Note that the SMB and HML portfolios are reconstituted annually, and the MOM portfolios are reconstituted monthly. The rebalancing frequencies of these common factors are inconsistent with the daily rebalancing of our DERL strategy. Consequently, while our investment scope includes the top 500 stocks in terms of market capitalization, the DERL portfolio has significantly positive loadings on the SMB factor. Nonetheless, the risk-adjusted daily returns ($\alpha$) of our DERL portfolio are above 0.03\%, or 7.5\% per annum, and are significant across different factor models, suggesting that these common factors cannot fully account for the portfolio returns.
Columns 3-5 tabulate the regression results for different subperiods. Consistent with the findings for the full sample, the DERL portfolio returns have significant loading on the market factor, and the risk-adjusted returns remain significant across different factor models. 
The coefficients of the market factor during different subperiods range from 0.60 to 0.65, all significant at the 1\% level. 

Finally, columns 6-7 present the estimation results under different volatility regimes. While the DERL portfolio has significant loadings on the market factor, the coefficient estimates are quite different across different volatility regimes. For instance, the coefficient of the market factor is around 0.75 when the market volatility is low, which drops to around 0.62 when the market volatility is high. This indicates that the DERL agent learns the timing ability to adjust its market exposure according to the market volatility conditions. The daily risk-adjusted returns $\alpha$  are 0.01\% and 0.05\%, or 2.5\%  or 12.5\% per annum under low and high volatility regimes respectively, and are both statistically significant.

\subsection{Portfolio Decision of DERL and Economic Insights}\label{sec:interpretibility}
Understanding how the DERL agent works is challenging due to its complex, layered, and nonlinear structure. In this section, we aim to analyze the decision patterns identified by the RL agent in DERL by linking the stock weights it produces to a set of standard stock characteristics. We focus on characteristics that capture stock-level liquidity (illiquidity \citep{amihud02}, bid-ask spread, share turnover, and number of no-trade days), recent price trends, and risk (return volatility, beta, and idiosyncratic volatility)\footnote{A brief description of their calculation methods can be found in \S \ref{ec:characteristics}.}. These characteristics are calculated using a rolling window method, with window sizes of 7, 14, or 30 calendar days, to capture the trading patterns of stocks over different time periods. 
The characteristics are then cross-sectionally standardized to have zero mean and unit variance. Considering the multicollinearity among the characteristics, we apply lasso regression period-by-period to select the most relevant characteristics for the stock weights\footnote{The stock weights are multiplied by 100 for ease of presentation.}. We then calculate the selection rates, reflecting how often each characteristic is chosen by the lasso algorithm, along with their time-series averages and corresponding $t$-values. 
Table \ref{tab: lasso} presents the main results.

\begin{table}[htbp]
\TABLE
{Lasso regression analysis of stock weights on standard characteristics \label{tab: lasso}}
{
\begin{tabular}{lcccccccccc}
\toprule
           &           \multicolumn{4}{c}{Liquidity}         & &       &  &    \multicolumn{3}{c}{Risk}  \\
  \cmidrule(r){2-5} \cmidrule(r){9-11}
  &     Illiq & Spread & Turn & Ztrade &            &  Trend &            &    Retvol &          Beta &  Ivol                     \\
\midrule
\multicolumn{8}{l}{Panel A: Full sample}\\
\midrule
$\%sel_{7d}$ &     35.56  &     43.06  &     40.21  &     37.03  &            &     62.08  &            &     52.52  &            &            \\

$\beta_{7d}$ & -0.0160*** &  0.0034*** &  0.0086*** & -0.0150*** &            & -0.0210*** &            & -0.0100*** &         &         \\

           &    [-3.99] &     [5.92] &     [7.35] &    [-7.13] &            &   [-38.50] &            &   [-18.42] &         &         \\

$\%sel_{14d}$ &     31.39  &     52.95  &     34.95  &     31.04  &            &     94.24  &            &     76.46  &         &         \\

$\beta_{14d}$ &  0.0385*** &  0.0231*** &  0.0076*** &   -0.0020  &            &  0.0741*** &            &  0.0505*** &         &         \\

           &     [6.27] &    [23.99] &     [3.67] &    [-0.47] &            &    [81.25] &            &    [51.51] &         &         \\

$\%sel_{30d}$ &     35.24  &     36.28  &     36.70  &     30.24  &            &     57.94  &            &     24.58  &     75.17  &     47.78  \\

$\beta_{30d}$ &   -0.0090  &  0.0028*** & -0.0090*** &  0.0168*** &            &  0.0047*** &            &  0.0292*** & -0.0320*** & -0.0100*** \\

           &    [-1.63] &     [5.31] &    [-6.79] &     [2.66] &            &    [15.58] &            &    [10.25] &   [-20.34] &    [-4.63] \\
\midrule
\multicolumn{8}{l}{Panel B1: Low volatility regime}\\
\midrule
 $\%sel_{7d}$ &     34.48  &     40.77  &     39.41  &     35.33  &            &     60.84  &            &     51.14  &            &            \\

$\beta_{7d}$ & -0.0160*** &  0.0038*** &  0.0066*** & -0.0190*** &            & -0.0220*** &            & -0.0100*** &         &         \\

           &    [-3.70] &     [6.48] &     [7.68] &    [-5.32] &            &   [-28.50] &            &   [-13.35] &         &         \\

$\%sel_{14d}$ &     32.36  &     53.82  &     33.99  &     30.54  &            &     95.50  &            &     77.61  &         &         \\

$\beta_{14d}$ &  0.0343*** &  0.0214*** &  0.0041*** &    0.0004  &            &  0.0740*** &            &  0.0488*** &         &         \\

           &     [3.58] &    [19.82] &     [3.70] &     [0.05] &            &    [61.10] &            &    [40.90] &         &         \\

$\%sel_{30d}$ &     33.86  &     35.55  &     34.26  &     28.07  &            &     56.36  &            &     23.06  &    74.63  &    44.31  \\

$\beta_{30d}$ &   -0.0140* &  0.0029*** & -0.0050*** &    0.0179* &            &  0.0040*** &            &  0.0242*** & -0.0250*** & -0.0110*** \\

           &    [-1.74] &     [5.34] &    [-7.90] &     [1.80] &            &    [12.41] &            &     [6.68] &   [-17.50] &    [-3.49] \\

\midrule
\multicolumn{8}{l}{Panel B2: High volatility regime}\\
\midrule

$\%sel_{7d}$ &     36.64  &     45.35  &     41.01  &     38.73  &            &     63.33  &            &     53.90  &            &            \\

$\beta_{7d}$ &  -0.0150** &  0.0031*** &  0.0106*** & -0.0120*** &            & -0.0210*** &            & -0.0090*** &         &         \\

           &    [-2.27] &     [3.17] &     [5.05] &    [-4.32] &            &   [-24.93] &            &   [-12.75] &         &         \\

$\%sel_{14d}$ &     30.42  &     52.08  &     35.92  &     31.55  &            &     92.98  &            &     75.31  &         &         \\

$\beta_{14d}$ &  0.0428*** &  0.0249*** &  0.0112*** &   -0.0040  &            &  0.0742*** &            &  0.0522*** &         &         \\

           &     [4.66] &    [14.79] &     [2.91] &    [-1.08] &            &    [49.45] &            &    [33.12] &         &         \\

$\%sel_{30d}$ &     36.61  &     37.01  &     39.13  &     32.40  &            &     59.53  &            &     26.11  &     75.72  &     51.25  \\

$\beta_{30d}$ &   -0.0050  &  0.0028*** & -0.0120*** &   0.0157** &            &  0.0055*** &            &  0.0342*** & -0.0400*** & -0.0090*** \\

           &    [-0.59] &     [2.88] &    [-4.80] &     [1.97] &            &    [10.21] &            &     [7.93] &   [-13.93] &    [-3.15] \\
\bottomrule
\end{tabular}
}
{\textit{Note}. This table tabulates the results of cross-sectional lasso regression of stock weights on standard characteristics. The characteristics are calculated using the rolling-window method, with window size being either 7, 14, or 30 calendar days. We report the selection rates of each characteristic, and the time-series average of the regression coefficient over all testing periods. 
The $t$-values with Newey-West adjustments are reported in brackets, and the asterisks *, **, and *** denote the 10\%, 5\%, and 1\% levels of statistical significance, respectively. }
\end{table}

For the full sample, Panel A of Table \ref{tab: lasso} shows that price trends are most likely to be chosen by the lasso algorithm. For instance, the selection rate for the 14-day price trend reaches 94\%, much higher than the other characteristics. Meanwhile, the selection rates for the 7- and 30-day trends are 62\% and 58\%, respectively. The time-series averages of the regression coefficients for the 7-, 14-, and 30-day price trends are -0.021, 0.074, and 0.005, respectively, all of which are statistically significant. This suggests that DERL favors stocks that have performed well over the past $14$ or $30$ days but have experienced a pullback in the last $7$ days.
Among characteristics related to firms' risk, return volatility calculated using past $7$ and $14$ days returns and the market beta estimated from the CAPM model have relatively large associations with the stock weights, with selection rates of $53\%$, $76\%$, and $75\%$, respectively. Moreover, the time-series averages of the regression coefficients are $-0.01$, $0.051$, and $-0.032$, respectively, all of which are significant. Therefore, DERL favors stocks with low systematic risk that have shown volatility over the past 14 days but have stabilized in the most recent $7$ days.
Finally, given that our investment universe contains the largest 500 stocks in the market, we find that liquidity characteristics are less relevant to the stock weights, with selection rates all below $50\%$.

Panels B1 and B2 present the results during the low and high volatility regimes, respectively. Consistent with the findings in Panel A, characteristics related to price trends and risks are most relevant to the stock weights chosen by the DERL agent. The associations between price trends and stock weights are generally similar under different market volatility conditions. The selection rates for 7- and 14-day price trends in low (high) volatility regimes are 61\% (95\%) and 63\% (93\%), respectively. The time-series averages of 7-day coefficients are significantly negative, while the averages of 14-day coefficients are significantly positive. This indicates that, during different volatility regimes, portfolio choices from DERL align with a ``7-day reversal and 14-day momentum" strategy. The associations between risk characteristics and stock weights vary across different market conditions. In particular, the selection rates of market beta during low and high volatility regimes are $75\%$ and $76\%$, respectively. The time-series average of its regression coefficient is $-0.025$ during low volatility and decreases to $-0.040$ when market volatility is high. Consistent with the findings in Table \ref{tab:factor_analysis}, these results indicate that the DERL agent has volatility timing capability and reduces investments in stocks with high systematic risks during periods of market stress.

\subsection{Ablation Study}\label{sec:ablation}

%\ins{Compare to standard embedding method}

%\ins{meta-learning}

The DERL framework incorporates three major deep learning methods, namely generative autoencoder, meta-learning, and reinforcement learning, to enhance the agents' ability to continuously learn and adjust their portfolios based on new data and market conditions. To evaluate the contribution of each component to model performance, we conduct a series of ablation studies, and the results are presented in Table \ref{tab: ablation study}.

Panel A shows the results for the full sample. We first replace the TD3 algorithm with two other RL algorithms: A2C \citep{mnih2016asynchronous} and DDPG \citep{lillicrap2015continuous}. All three RL algorithms within the DERL framework outperform versions without dynamic updating and versions without both dynamic updating and embeddings. This indicates the compatibility of different RL algorithms with the DERL framework. While replacing the TD3 RL algorithm with either A2C or DDPG results in lower Sharpe and Sortino ratios, the difference between the TD3 and A2C models is not statistically significant. The significant outperformance of TD3 over DDPG can be attributed to TD3's improved training stability and algorithmic superiority over DDPG.

After removing the dynamic learning feature (Meta-learning) from the baseline model, the agent generates returns with lower means and significantly higher standard deviations. Consequently, the Sharpe (Sortino) ratio drops significantly from 1.04 (1.62) to 0.64 (0.90). When both the embedding and meta-learning features are removed, the agent performs even worse, particularly in managing portfolio risks, producing realized returns with a standard deviation of 0.23, compared to only 0.14 in the full model. As a result, the Sharpe and Sortino ratios of the model decrease to approximately 0.50 and 0.69, respectively, after the removal of these two critical features.

Panels B1 and B2 present the results of the ablation study under different market volatility conditions. Overall, the outperformance of the agent with full components in DERL compared to other alternative agents is not significant, except for the A2C model. However, when market volatility is high, the DERL agent yields superior performance. Our ablation study reveals the crucial role of dynamic embedding in managing portfolio risks and enhancing portfolio performance, especially during periods of market stress.

We also find that the agent using embeddings of the current state outperforms the agent without embeddings, as observed from the results in Lines 5 and 6 of each panel, showing the importance of low-dimensional embeddings for noise reduction. Additionally, the performance of the agent that encodes the next state surpasses that of the agent encoding the current state, with statistical significance. This advantage is even more pronounced in high-volatility regimes, suggesting that next-state embeddings provide more accurate and informed latent states for portfolio allocation, particularly during periods of market stress.
 
\begin{table}[htbp]
\TABLE
{Ablation study \label{tab: ablation study}}
{\begin{tabular}{cccccccccccc}
\toprule
\multicolumn{3}{c}{Model specification}  && \multicolumn{4}{c}{Return performance} && \multicolumn{3}{c}{FF3 factor analysis}\\
\cmidrule(r){1-3} \cmidrule(r){5-8} \cmidrule(r){10-12}
Embedding & Meta & RL &            &       Mean &        STD &              SR &         ST  & & $\alpha$  &Market &Adj. $R^2$\\
\midrule
           \multicolumn{8}{l}{Panel A: Full samples (N=7550)}\\
\midrule
next state &        yes &        TD3 &            &    0.1481  &    0.1423  &    1.0407  &    1.6200  &            &  0.0003*** &  0.6380*** &    0.7467  \\

next state &        yes &        A2C &            &    0.1329  &    0.1424  &    0.9334  &    1.4212  &            &  0.0003*** &  0.5810*** &    0.6018  \\

 next state  &   yes      &    DDPG                  &            &    0.1239  &    0.1450  &   0.8544** &  1.2296*** &            &  0.0002*** &  0.7074*** &    0.8614  \\

next state &         no &        TD3 &            &    0.1135  &    0.1775  &  0.6394*** &  0.9018*** &            &   0.0001** &  0.8392*** &    0.8175  \\

current state &        yes &        TD3 &            &    0.1238  &    0.1681  &  0.7361*** &  1.0557*** &            &  0.0002*** &  0.8161*** &    0.8603  \\

        no &         no &        TD3 &            &    0.1158  &    0.2328  &  0.4975*** &  0.6934*** &            &    0.0000  &  1.1201*** &    0.8500  \\

\midrule
\multicolumn{8}{l}{Panel B1: Low volatility regime (VIX<17.91, N=3371)}\\
\midrule

next state &        yes &        TD3 &            &    0.2369  &    0.0841  &    2.8171  &    4.6230  &            &   0.0001** &  0.7499*** &    0.7954  \\

next state &        yes &        A2C &            &    0.2385  &    0.1002  &   2.3795** &   3.9107** &            &    0.0001  &  0.7317*** &    0.5098  \\

next state           &     yes        &   DDPG         &            &    0.2568  &    0.0897  &    2.8640  &    4.6795  &            &   0.0001** &  0.8028*** &    0.7675  \\

next state &         no &        TD3 &            &    0.2800  &    0.1060  &    2.6429  &    4.2363* &            &    0.0001  &  0.9277*** &    0.7460  \\

current state &        yes &        TD3 &            &    0.2805  &    0.1013  &    2.7689  &    4.5071  &            &    0.0001* &  0.9083*** &    0.7778  \\

        no &         no &        TD3 &            &    0.3458  &    0.1280  &    2.7029  &    4.3396  &            &    0.0000  &  1.1683*** &    0.8157  \\

\midrule
\multicolumn{8}{l}{Panel B2: High volatility regime (VIX$\geq$ 17.91, N=3375)}\\
\midrule

next state &        yes &        TD3 &            &    0.0589  &    0.1827  &    0.3223  &    0.4960  &            &  0.0005*** &  0.6198*** &    0.7424  \\

next state &        yes &        A2C &            &    0.0271  &    0.1744  &    0.1554  &    0.2310  &            &  0.0003*** &  0.5573*** &    0.6398  \\

  next state         &    yes         &    DDPG        &            &   -0.0090  &    0.1841  & -0.0490*** & -0.0685*** &            &  0.0002*** &  0.6920*** &    0.8875  \\

next state &         no &        TD3 &            &   -0.0538  &    0.2271  & -0.2370*** & -0.3256*** &            &    0.0001  &  0.8241*** &    0.8357  \\

current state &        yes &        TD3 &            &   -0.0333  &    0.2147  & -0.1550*** & -0.2165*** &            &   0.0002** &  0.8007*** &    0.8817  \\

        no &         no &        TD3 &            &   -0.1148  &    0.3028  & -0.3791*** & -0.5160*** &            &    0.0000  &  1.1102*** &    0.8576  \\

\bottomrule           
\end{tabular}
}
{\textit{Note}. This table presents the results of the ablation study to examine the contribution of each of the three components of our framework, namely, the embedding model, meta-learning (Meta), and reinforcement learning (RL) to the model performance. Panels A, B1, and B2 present the result for the full sample, the low volatility subsample, and the high volatility subsample, respectively. The asterisks *, **, and *** denote statistical significance at the 10\%, 5\%, and 1\% levels, respectively, for the null hypothesis that the full baseline model underperforms the alternative model.}
\end{table}

Table \ref{tab: ablation study} demonstrates that the embedding and meta-learning components in our DERL framework are crucial for enhancing model performance. To evaluate the role of the embedding component in managing noisy data, we conduct the following time-series regression:
\begin{equation}
\mathrm{EMB}_{t} = b_0+b_1 \mathrm{Market}_{t} + b_2 \mathrm{VIX}_t +u_t,
\end{equation}
where $\mathrm{EMB}_t$ represents the embedding contribution, defined as the difference in returns between the fourth and sixth models listed in Panel A of \ref{tab: ablation study}, and $u_t$ is the residual term. The regression incorporates two market variables: market return and VIX, a widely-used proxy for market uncertainty. Columns 2-4 of Table \ref{tab: ablation effect} display the estimation results. Consistent with our objective for deploying embedding, models (1) and (2) reveal that the embedding contribution is more pronounced when the market return is low or the VIX is high, indicating market frictions. When both market variables are included, model (3) shows that the coefficient of market return remains significantly negative, and the VIX coefficient is significantly positive, although the level of significance decreases.

\begin{table}[htbp]
\setlength{\tabcolsep}{6pt}
\TABLE
{Component contributions and market conditions \label{tab: ablation effect}}
{\begin{tabular}{lcccccccccc}
\toprule
& &\multicolumn{3}{c}{Embedding}  && && \multicolumn{3}{c}{Meta-learning} \\
\cmidrule(r){3-5} \cmidrule(r){9-11} 
 & & (1) & (2) & (3)           &        &  &&       (1) &              (2) &         (3) \\
\toprule

 Intercept &            &  0.0001*** & -0.0013*** &   -0.0001  &            &  Intercept &            &   -0.0001  &   -0.0003* &   -0.0003* \\

           &            &     [3.49] &    [-5.35] &    [-1.00] &            &            &            &    [-0.48] &    [-1.72] &    [-1.69] \\

    Market &            & -0.2790*** &            & -0.2779*** &            &  
    $\mathrm{DMkt}$ &            &    0.0003  &            &    0.0001  \\

           &            &   [-39.47] &            &   [-53.83] &            &            &            &     [1.55] &            &     [0.46] \\

       VIX &            &            &  0.0063*** &    0.0012* &            &  
       $\mathrm{DVIX}$ &            &            &   0.0003** &   0.0003** \\

           &            &            &     [4.84] &     [1.76] &            &            &            &            &     [2.49] &     [2.37] \\

   Adjusted $R^2$        &            &    0.6073  &    0.0150  &    0.6077  &            &     &            &    0.0020  &    0.0047  &    0.0047  \\

\bottomrule           
\end{tabular}
}
{\textit{Note}. This table presents time-series regression of the contributions of embedding and meta learning on market variables. The $t$-values with Newey-West adjustments are reported  in brackets, and the asterisks *, **, and *** denote the 10\%, 5\%, and 1\% levels of statistical significance, respectively.}
\end{table}

Similarly, let $\mathrm{Meta}_t$ represent the meta-learning contribution, calculated as the difference in returns between the first and fourth models listed in Panel A of \ref{tab: ablation study}. Since DERL applies meta-learning every 42 days, $\mathrm{Meta}_t$ is expected to be insignificant for the first 42 days of each training segment. The corresponding $t$ statistic is $-0.22$, aligning with our expectation. For the remaining samples in each segment, the $t$ statistic for $\mathrm{Meta}_t$ is $2.04$, indicating that meta-learning significantly enhances portfolio performance.

We then run the following time-series regression to explore how meta-learning contributes to managing nonstationarity in the market:
\begin{equation}
    \mathrm{Meta}_{t} = b_0+b_1 \mathrm{DMkt_{t}} + b_2 \mathrm{DVIX_{t}} +u_t.
\end{equation}
We construct two variables, $\mathrm{DMkt}$ and $\mathrm{DVIX}$, to proxy potential structural changes in market conditions, defined as
\begin{equation}
 \begin{aligned}
   \mathrm{DMkt_{t}} &= \left | \frac{\mathrm{Market}_{t}-\mu_{\mathrm{Market}}} {\sigma_{\mathrm{Market}}} \right|, \\
   \mathrm{DVIX_{t}} &= \left | \frac{\mathrm{VIX}_{t}-\mu_{\mathrm{VIX}}} {\sigma_{\mathrm{VIX}}} \right|,
\end{aligned}   
\end{equation}
where $\mu_{\mathrm{Market}}$ and $\mu_{\mathrm{VIX}}$ are the unconditional means, and $\sigma_{\mathrm{Market}}$ and $\sigma_{\mathrm{VIX}}$ are the unconditional standard deviations of market returns and VIX over a 3-year training period.

Columns 6-8 of Table \ref{tab: ablation effect} present the estimation results of the time-series regressions. The results show that deviations in current market returns from their historical distribution do not significantly impact the contribution of meta-learning. However, when market volatility patterns shift, the estimation results from models (1) and (3) indicate that meta-learning significantly enhances model performance.

\begin{comment}

\subsection{Model Robustness}

\subsubsection*{Compare to Other RL algorithms}

\ins{Compare to A2C, DDPG, reviewer 1}
% The annualized returns of S\&P 500 and CSI 500 are shown in Table \ref{tab: SP500 latent space} and Table \ref{tab: CSI500 latent space}. In both markets, investment performance exhibits an inverted U-shape with latent size. Returns first rise as the latent size gets smaller, peaking when the size reaches around 300-600 and then dropping again. These results are intuitive, as the rise in the inverted U-shape gets the benefits of sample efficiency from embedding, while the lack of information encoded by a small embedding layer causes the downward trend of this U-shape. In addition, Table \ref{tab: SP500 latent space} and Table \ref{tab: CSI500 latent space} also show that dynamically updating the embedding can improve the annualized return performance by about 2\%-13\%.
\end{comment}

\section{Conclusion}\label{sec: conclusion}
This paper introduces a reinforcement learning (RL) framework for dynamic portfolio allocation that enhances both sample efficiency and risk management. By embedding the original market state into a more representative latent space prior to learning the strategy, we improve the handling of high-dimensional and noisy financial data. We use generative autoencoders to project states into the embedding space and incorporate the fully online meta-learning techniques to dynamically adapt the embedding encoder, capturing the most recent financial transition patterns.

Our dynamic portfolio allocation framework introduces novel advancements in portfolio management. Firstly, the RL agent within our framework is designed to automatically handle portfolio allocation in an end-to-end manner, adeptly navigating high-dimensional, low signal-to-noise, and non-stationary market conditions over long trading horizons. Secondly, it enhances market information management by employing an embedding method that captures and converts essential features of high-dimensional financial data into a more manageable lower-dimensional representation. This is particularly advantageous for rapidly identifying and responding to key market trends, and is especially effective in adapting to non-stationary market shifts. Thirdly, our framework significantly improves risk management by dynamically updating its understanding of market transitions. This allows for the quick identification of potential risks and the implementation of preemptive measures to mitigate them. Such capabilities ensure swift reactions to market changes, thereby minimizing potential risks and maximizing profits.

An empirical study based on the top 500 market-value stocks in each subperiod of the U.S. stock market demonstrates that our framework outperforms the two-step predict-then-optimize model, as well as the value- and equal-weighted strategies. Moreover, the superior performance of our DERL framework cannot be fully explained by well-known factor models, such as the \cite{FAMA19933} three-factor model and the \cite{CAR1997} four-factor model, and is even more pronounced during periods of market stress. Further analysis reveals that the DERL agent has volatility timing capability, reducing the portfolio's exposure to the market during periods of market stress. Through a series of ablation studies, we find that the framework's performance is robust across different RL algorithms. Additionally, embedding and meta-learning effectively address the challenges posed by noisy and non-stationary data, making them crucial for improving portfolio performance.

\bibliographystyle{informs2014}
\bibliography{ref}

\clearpage
\renewcommand{\thepage}{ec\arabic{page}}  

\begin{APPENDICES}
 
\renewcommand{\thetable}{EC.\arabic{table}}   
\renewcommand{\thefigure}{EC.\arabic{figure}}
\renewcommand{\theequation}{EC.\arabic{equation}}
\renewcommand{\thetheorem}{EC.\arabic{theorem}}
\renewcommand{\theproposition}{EC.\arabic{proposition}}
\renewcommand{\thelemma}{EC.\arabic{lemma}}
\renewcommand{\thecorollary}{EC.\arabic{corollary}}

\counterwithin{equation}{section}
\counterwithin{table}{section}
\counterwithin{figure}{section}
\counterwithin{lemma}{section}
\setcounter{equation}{0}
\setcounter{page}{1}
\newcounter{partsection}
\setcounter{partsection}{1}
\renewcommand{\thesection}{EC.\arabic{partsection}}

    \vspace*{0.1cm}
   \begin{center}
      \large\textbf{E-Companion --- Reinforcement-Learning Portfolio Allocation with Dynamic Embedding of Market Information}\\
   \end{center}
   \vspace*{0.3cm}

\section{Further Related Literature}\label{sec:related works} 

\subsubsection*{Machine Learning in Financial Applications}\label{sec:review on ML based algorithmic trading}

Machine learning (ML) is a category of data-driven algorithms that utilize statistical models to identify patterns and generate predictions based on historical data. Financial data is well suited for ML techniques, and as a result, a significant body of research has applied ML to finance for various purposes \citep{cong2020alphaportfolio, huang2022machine}. Such techniques can analyze large volumes of data, identify non-linear patterns, and generate insights that traditional models may overlook. ML-assisted techniques have been widely applied in finance to identify patterns in data and improve investment performance \citep{ban2018machine,KX2023, jiang2023re}. 

In particular, \cite{gu2020empirical} and \cite{FNW2020} find that the nonlinear integration of large dimensional firm characteristics using machine learning methods helps to improve the predictability of cross-section asset returns, and the long-short portfolios based on the learning signals can produce superior
out-of-sample performance. 
\cite{cong2021deep} provides a deep sequence model for asset pricing, highlighting its ability to deal with high-dimensional, nonlinear, interactive, and dynamic financial data. They show that Long short-term memory with an attention mechanism outperforms other models in portfolio performance. Furthermore, \cite{bryzgalova2023asset} used an ML-assisted factor analysis approach to estimate latent asset price factors using cross-sectional and time series targets. They show that this method leads to higher Sharpe ratios and lower pricing errors than conventional approaches when tested on a large-scale set of assets.

\subsubsection*{Reinforcement Learning}\label{sec:review on RL}

\begin{comment}
    In recent years, RL has been applied successfully to a wide range of domains, including recommendation systems \citep{zheng2018drn}, games \citep{silver2016mastering}, and robotics \citep{kober2013reinforcement}. 
\end{comment}

Reinforcement learning (RL) has gained significant attention in recent years due to its ability to make decisions based on real-time feedback. RL is a subfield of machine learning that is concerned with how an agent can learn to make decisions by interacting with an environment, where the agent receives feedback in the form of rewards or penalties for each action taken \citep{gosavi2009reinforcement}. RL has been used in various applications, including gaming \citep{silver2017mastering}, communication network \citep{qu2022scalable}, marketing \citep{wang2022deep, liu2022dynamic}, and finance \citep{cong2020alphaportfolio}. 

One notable achievement in RL is AlphaGo \citep{silver2017mastering}, which defeated the top human players in the game of Go. The success of AlphaGo has led to increased interest in RL, including its application to finance. \cite{hambly2021recent} review recent advances in RL in finance. RL has the potential to improve trading decision-making by finding optimal strategies, reducing error propagation, and capturing the complexity of financial markets. \cite{liu2021finrl} introduces the FinRL library as a unified framework for deploying deep reinforcement learning algorithms in quantitative finance.
However, efficiently implementing RL for portfolio management and trading is currently hindered by two main challenges: excessive noise and high dimensionality. Noise in stock market data can lead to overestimation of value functions and suboptimal policies, while high dimensionality can make it difficult for RL agents to achieve optimal portfolio diversification.

 The seminal work by \cite{cong2020alphaportfolio} proposed a deep RL framework that outperforms traditional portfolio management methods. Our approach is distinct in that we propose a different framework, a dynamic embedding reinforcement learning framework. This framework combines three major deep learning methods: reinforcement learning for profitable trading decisions, generative models for summarizing and analyzing high-dimensional stock market information, and meta-learning for adapting the trading framework to changing market conditions. By utilizing these methods together, we provide an end-to-end solution for portfolio allocation.

\subsubsection*{Embedding Methods}

Embedding methods in finance are important because they enable the transformation of complex financial data into continuous vector representations that facilitate advanced machine learning and predictive analytics. These embeddings help in capturing underlying patterns, relationships, and trends in financial data, thereby improving the accuracy of models used for tasks such as risk assessment and portfolio allocation. \cite{kelly2019characteristics} present instrumented PCA (IPCA), a technique that utilizes covariates for dimensionality reduction but is limited by its reliance on linear models. On the other hand, \cite{gu2021autoencoder} applies autoencoder neural networks for unsupervised dimension reduction, incorporating both covariate information and returns. This method compresses returns into a low-dimensional space, allowing for nonlinear and interactive effects of stock characteristic covariates on factor exposures.

In addition to conventional embedding methods, generative models learn latent representations of underlying data to generate meaningful content, resulting in more effective and insightful embeddings. Generative models are machine learning models that create new data similar to the training set. In financial applications, \cite{chen2023deep} introduces a deep neural network asset pricing model for individual stock returns, integrating extensive conditioning information and accounting for time variation through recurrent neural networks and generative adversarial networks. Similarly, \cite{cong2019textual} proposes a method for analyzing large-scale text data by combining neural network language processing with generative statistical modeling. In our framework, we utilize generative encoders to summarize and analyze high-dimensional stock market information.

\subsubsection*{Meta-Learning}

% Meta-learning and generative models are two techniques that have gained traction in artificial intelligence. 
Meta-learning is concerned with learning how to learn from a set of related tasks and has been applied in various domains \citep{bastani2021predicting, anderer2022adaptive, bastani2022meta}. Our work is motivated by the fully online meta-learning framework \cite{rajasegaran2022fully}, which allows the model to adapt continuously across changing tasks and input distributions without resetting the model. This is particularly crucial in finance, where the financial market is constantly changing, and any static strategy may not adequately adapt to the non-stationary market.

There is sparse application of meta-learning in financial applications, and being a relatively new technique, it is an area that requires more exploration \citep{hambly2021recent}. \cite{bastani2022meta} proposes a meta-dynamic pricing algorithm that learns the unknown demand parameters shared across a sequence of dynamic pricing experiments for related products. They balance meta-exploration and meta-exploitation and account for uncertainty in the estimated prior, demonstrating its effectiveness through numerical experiments on synthetic and real auto loan data. \cite{duan2022target} proposed a transfer learning approach that incorporates market-specific knowledge into a target market to improve the performance of machine learning algorithms.

\section{Details of Reinforcement Learning and TD3 algorithm}\label{ec:details of RL}
Reinforcement learning (RL) is a framework to teach machines to make decisions based on rewards received from an environment. In model-based RL, an agent interacts with an environment modeled as a Markov decision process (MDP), denoted as $M=\{\mathcal{S}, \mathcal{A}, \mathbb{P}, r, \gamma\}$. Here, $\mathcal{S}$ is the set of possible states in the environment. A particular state $s \in S$ represents the current conditions of the system, such as the current holdings of all equities and market information. $\mathcal{A}$ is the set of feasible actions, which represent the decisions that the agent can make. In finance, an action $a \in \mathcal{A}$ could correspond to buying or selling a certain amount of a particular asset.

The transition probability $\mathbb{P}(\boldsymbol{s}^{'} |\boldsymbol{s}, \boldsymbol{a})$ represents the probability of transitioning to a new state $\boldsymbol{s}'$ when taking action $\boldsymbol{a}$ in the current state $\boldsymbol{s}$. After taking action $\boldsymbol{a}_{t}$ in the $t$-th step, the agent receives a reward $r_t$ based on the path tuple $(\boldsymbol{s}_{t}, \boldsymbol{a}_{t}, \boldsymbol{s}_{t+1})$, where $r_{t}=r\left(\boldsymbol{s}_{t}, \boldsymbol{a}_{t}, \boldsymbol{s}_{t+1}\right)$. The discount factor $\gamma$ balances the importance of immediate and future rewards. In RL, the cumulative reward is the discounted sum of the rewards of each step over a period $T$ (which can be infinite): $\sum_{t=1}^{T} \gamma^{t-1} r_{t}$. The objective of RL is to learn a policy $\pi$ that maximizes the expected cumulative reward in the MDP framework.
\subsection{Bellman Equation and Bellman Optimality}

Given a fixed policy $\pi$, we can determine the state transition dynamics:
\begin{equation}
    \mathbb{P}^\pi(\boldsymbol{s}'|\boldsymbol{s}) = \int_{\boldsymbol{a}\in\mathcal{A}(\boldsymbol{s})}\pi(\boldsymbol{a}|\boldsymbol{s}) \mathbb{P}(\boldsymbol{s}'|\boldsymbol{s},\boldsymbol{a}) \mathrm{d}  \boldsymbol{a}.
\end{equation}
With the state transition dynamics $\mathbb{P}^\pi(\boldsymbol{s}'|\boldsymbol{s})$, we can calculate the probability of any trajectory $\boldsymbol{\tau}^\pi(\boldsymbol{s}_0,\boldsymbol{a}_0,\boldsymbol{s}_1\cdots,\boldsymbol{s_T})$. By taking the expectation over all trajectories, we can estimate the expected sum of discounted future returns. We define the value function $V^{\pi}_t(\boldsymbol{s}): \Pi\times S\times [T]\rightarrow\mathbb{R}$ as the expected cumulative discounted return when visiting state $\boldsymbol{s}$ at time $t\leq T$:
\begin{equation}
V^{\pi}_t(\boldsymbol{s})=\mathbb{E}_{\boldsymbol{\tau}^\pi}\left[\sum_{k=t}^{T} \gamma^{k-t} r_{k}  \mid \boldsymbol{s}_{t}=\boldsymbol{s}\right].
\end{equation}
For example, if we take $\gamma=\frac{1}{1+\delta_f}$ as the discount rate for the value function, the value function may represent the net present value (NPV) of the portfolio at time $t$. 
For notation simplicity, we abbreviate $\mathbb{E}_{\boldsymbol{\tau}^{\pi}}$ as $\mathbb{E}_{\pi}$ for the rest of the paper.
% With the transitions and rewards, we can derive the following value function
% \begin{equation}
%     V^{\pi}_t(\boldsymbol{s})=\mathbb{E}_{\pi}\left[\sum_{k=t}^{T} (\frac{1}{1+\delta_f})^{k-t} \left[\left(b_{t+k+1}+\boldsymbol{1}^\top\boldsymbol{x}_{t+k+1} \right)-\left(b_{t+k}+\boldsymbol{1}^\top\boldsymbol{x}_{t+k}\right)\right]\mid \boldsymbol{s}_{t}=\boldsymbol{s} \right].
% \end{equation}

To account for the stochasticity of the policy  $\pi(\boldsymbol{a}|\boldsymbol{s})$, we introduce the state-action value function $Q_t^\pi(s_t,a_t):\Pi\times\mathcal{S}\times\mathcal{A}(s)\times[T]\rightarrow \mathbb{R}$. This function allows us to estimate the expected value from taking a specific action $\boldsymbol{a}_t$ in state $\boldsymbol{s}_t$ at time $t$ under policy $\pi$. We have
\begin{align}\label{eq:Q from V}
Q_t^\pi(\boldsymbol{s}_t,\boldsymbol{a}_t)&=\mathbb{E}_{\boldsymbol{s}_{t+1}\sim \mathbb{P}^\pi(\boldsymbol{s}'|\boldsymbol{s})}\left[r\left(\boldsymbol{s}_{t}, \boldsymbol{a}_{t}, \boldsymbol{s}_{t+1}\right)+\frac{1}{1+\delta_f}V_{t+1}^\pi(\boldsymbol{s}_{t+1})\right].
\end{align}
This state-action function $Q^\pi$ and state value function $V^\pi$ can be derived from each other through equation (\ref{eq:Q from V}) and 
\begin{equation}
    V_t(\boldsymbol{s}_t) = \int_{\boldsymbol{a}\in \mathcal{A}(\boldsymbol{s}_t)} \pi(\boldsymbol{a}_t|\boldsymbol{s}_t)\mathbb{P}(\boldsymbol{s}_{t+1}|\boldsymbol{s}_t,\boldsymbol{a}_t)Q^\pi(\boldsymbol{s}_t,\boldsymbol{a}_t) \text{d}\boldsymbol{a}_t = \mathbb{E}_\pi\left[Q^\pi(\boldsymbol{s}_t,\boldsymbol{a}_t)\right].
\end{equation}

The aim of reinforcement learning (RL) is to find the optimal policy $\pi^\star(\boldsymbol{a}|\boldsymbol{s})$ that maximizes the expected value function for any $\boldsymbol{s}$, which can be represented as follows: $\forall \boldsymbol{s}$ and $\boldsymbol{a}$, 
\begin{equation}
\pi^{\star}=\arg \max _{\pi \in \Pi} V^{\pi}(\boldsymbol{s})=\arg \max _{\pi \in \Pi} Q^{\pi}(\boldsymbol{s}, \boldsymbol{a}).
\end{equation}
In portfolio management, the optimal policy $\pi^\star$ can be interpreted as the agent's strategy for making purchase-sell decisions for each asset in the portfolio to maximize the expected net present value of their portfolio when observing the market state. The Bellman Optimality Equation \citep{sutton1998introduction} guarantees the existence of such an optimal policy:
\begin{proposition}[Bellman Optimality]
Let $\Pi$ denote the set of all non-stationary and randomized policies, and assume an infinite time horizon. Define:
$$
\begin{aligned}
V^{\star}(s) :=\sup _{\pi \in \Pi} V^\pi(s), \quad Q^{\star}(s, a) :=\sup _{\pi \in \Pi} Q^\pi(s, a) .
\end{aligned}
$$
There exists a stationary policy $\pi^\star$ such that for all $s \in \mathcal{S}$ and $a \in \mathcal{A}$,
$$
\begin{aligned}
V^{\pi^\star(s)} =V^{\star}(s),\quad Q^{\pi^\star}(s, a) =Q^{\star}(s, a) .
\end{aligned}
$$
The optimal value function satisfies 
\begin{equation}\label{eq:Bellman Eq}
Q_t^{\pi^\star}(\boldsymbol{s}_t,\boldsymbol{a}_t)=\mathbb{E}_{\boldsymbol{s}_{t+1}\sim \mathbb{P}^{\pi^\star}(\boldsymbol{s}'|\boldsymbol{s})}\left[r\left(\boldsymbol{s}_{t}, \boldsymbol{a}_{t}, \boldsymbol{s}_{t+1}\right)+\frac{1}{1+\delta_f} \left[\max_{\boldsymbol{a}_{t+1}\in \mathcal{A}(\boldsymbol{s}_{t+1})} Q_t^{\pi^\star}(\boldsymbol{s}_{t+1},\boldsymbol{a}_{t+1})\right]\right].
\end{equation}
\end{proposition}

The Bellman optimality equation demonstrates the existence of optimal policies and provides a approach to find $\pi^{\star}$ by finding the optimal value functions $Q^{\pi^{\star}}(\boldsymbol{s}, \boldsymbol{a})$ or $V^{\pi^{\star}}(\boldsymbol{s})$. Equation (\ref{eq:Bellman Eq}) is particularly useful in practice since it allows us to easily derive and evaluate policy without modeling transition dynamics, and can handle both discrete and continuous action spaces. This approach is especially important in financial market settings, where transition dynamics are difficult to capture and decisions are continuous.

The concept of modeling the value function directly, rather than attempting to estimate the transition dynamics $\mathbb{P}(\boldsymbol{s}'|\boldsymbol{s},\boldsymbol{a})$ forms the basis of  \textit{model-free} RL algorithms \citep{silver2014ddpg, silver2016mastering, fujimoto2018td3}. These algorithms use various function types and techniques to approximate the value function induced by a given policy, and seek to identify the value function that satisfies the equilibrium in equation \eqref{eq:Bellman Eq}. Our framework also employs model-free RL algorithms as the agents, given the difficulty of directly modeling the transition dynamics in financial markets.

\begin{lemma}\label{Bellman equation}
(Bellman Equation) If the state space and action space is discrete and finite, for a fixed policy $\pi$, we have the following recursive relation for the state value function $V^{\pi}(\boldsymbol{s})$ and the state-action value function $Q^{\pi}(\boldsymbol{s}, \boldsymbol{a})$
\begin{equation}
\begin{aligned}
V^{\pi}(\boldsymbol{s}) &=\sum_{\boldsymbol{a} \in \mathcal{A}(\boldsymbol{s})} \pi(\boldsymbol{a}|\boldsymbol{s}) \sum_{\boldsymbol{s}' \in \mathcal{S}} \mathbb{P}(\boldsymbol{s}'|\boldsymbol{s},\boldsymbol{a})\left[r(\boldsymbol{s, \boldsymbol{a}, \boldsymbol{s}')}+\gamma V^{\pi}\left(\boldsymbol{s}'\right)\right]=\sum_{\boldsymbol{a} \in \mathcal{A}(\boldsymbol{s})} \pi(\boldsymbol{a} \mid \boldsymbol{s}) Q^{\pi}(\boldsymbol{s}, \boldsymbol{a}).
\end{aligned}
\end{equation}
Similarly, we can also express the $Q(\boldsymbol{s}, \boldsymbol{a})$ as:
\begin{equation}
\begin{aligned}
Q^{\pi}(\boldsymbol{s}, \boldsymbol{a}) =\sum_{\boldsymbol{s}'\in\mathcal{S}} \mathbb{P}(\boldsymbol{s}'|\boldsymbol{s},\boldsymbol{a}) r_{t}(\boldsymbol{s}, \boldsymbol{a}, \boldsymbol{s}')+\gamma \sum_{\boldsymbol{s}' \in S} \mathbb{P}(\boldsymbol{s}'|\boldsymbol{s},\boldsymbol{a})V^{\pi}\left(\boldsymbol{s}'\right)
\end{aligned}
\end{equation}
\end{lemma}

\proof{Proof: }
To show a brief proof of the Bellman equation, we start with the definition of the state value function $V^{\pi}(\boldsymbol{s})$ and the state-action value function $Q^{\pi}(\boldsymbol{s}, \boldsymbol{a})$ under a fixed policy $\pi$:
\begin{align*}
V^{\pi}(\boldsymbol{s}) &= \mathbb{E}_{\pi}\left[\sum_{t=0}^{\infty} \gamma^t r_{t+1} \mid  \boldsymbol{s}_0 =  \boldsymbol{s}\right] \\
Q^{\pi}( \boldsymbol{s},  \boldsymbol{a}) &= \mathbb{E}_{\pi}\left[\sum_{t=0}^{\infty} \gamma^t r_{t+1} \mid \boldsymbol{s}_0 =  \boldsymbol{s},  \boldsymbol{a}_0 =  \boldsymbol{a}\right],
\end{align*}
where $r_{t+1}$ is the reward obtained at time $t+1$ and $\gamma$ is the discount factor.

Now, we can use the law of total expectation to expand these expressions as follows:
\begin{align*}
V^{\pi}(\boldsymbol{s}) &= \mathbb{E}_{\pi}\left[\sum_{t=0}^{\infty} \gamma^t r_{t+1} \mid \boldsymbol{s}_0 = \boldsymbol{s}\right] \\
&= \sum_{\boldsymbol{a} \in \mathcal{A}(\boldsymbol{s})} \pi(\boldsymbol{a}|\boldsymbol{s}) \mathbb{E}_{\pi}\left[\sum_{t=0}^{\infty} \gamma^t r_{t+1} \mid \boldsymbol{s}_0 = \boldsymbol{s}, \boldsymbol{a}_0 = \boldsymbol{a}\right] \\
&= \sum_{\boldsymbol{a} \in \mathcal{A}(s)} \pi(\boldsymbol{a}|\boldsymbol{s}) Q^{\pi}(s, a),
\end{align*}
where $\mathcal{A}(\boldsymbol{s})$ is the set of actions available in state $\boldsymbol{s}$.
Similarly, we can expand $Q^{\pi}(\boldsymbol{s}, \boldsymbol{a})$ as follows:
\begin{align*}
Q^{\pi}(\boldsymbol{s}, \boldsymbol{a}) &= \mathbb{E}_{\pi}\left[\sum_{t=0}^{\infty} \gamma^t r_{t+1} \mid \boldsymbol{s}_0 = \boldsymbol{s}, \boldsymbol{a}_0 = \boldsymbol{a}\right] \\
&= \mathbb{E}_{\pi}\left[r_{1} + \sum_{t=1}^{\infty} \gamma^t r_{t+1} \mid \boldsymbol{s}_0 = \boldsymbol{s}, \boldsymbol{a}_0 = \boldsymbol{a}\right] \\
&=\sum_{\boldsymbol{s}'\in\mathcal{S}} \mathbb{P}(\boldsymbol{s}'|\boldsymbol{s},\boldsymbol{a}) r_{t}(\boldsymbol{s}, \boldsymbol{a}) + \gamma \sum_{\boldsymbol{s}' \in S} \mathbb{P}(\boldsymbol{s}'|\boldsymbol{s},\boldsymbol{a})\mathbb{E}_{\pi}\left[\sum_{t=0}^{\infty} \gamma^t r_{t+2} \mid \boldsymbol{s}_1 = \boldsymbol{s}', \boldsymbol{a}1 = \boldsymbol{a}'\right] \\
&= \sum_{\boldsymbol{s}' \in S} \mathbb{P}(\boldsymbol{s}'|\boldsymbol{s},\boldsymbol{a}) r_{t}(\boldsymbol{s}, \boldsymbol{a}) + \gamma \sum_{\boldsymbol{s}' \in S} \mathbb{P}(\boldsymbol{s}'|\boldsymbol{s},\boldsymbol{a}) V^{\pi}(\boldsymbol{s}'),
\end{align*}
where $\boldsymbol{a}'$ is an action chosen by the policy $\pi$ in state $\boldsymbol{s}'$.

Therefore, we have proved the Bellman equation for the state value function $V^{\pi}(\boldsymbol{s})$ and the state-action value function $Q^{\pi}(\boldsymbol{s}, \boldsymbol{a})$ under a fixed policy $\pi$. \Halmos
\endproof

For small-scale problems, such as a tabular MDP, if a model $M=\{\mathcal{S}, \mathcal{A}, \mathbb{P}, r, \gamma\}$ is provided, we can directly compute the value function $V^{\pi}$ and the action-value function $Q^{\pi}$ based on the transition probabilities and expected reward dynamics. However,  for larger problems with continuous state and action spaces, such as those encountered in finance settings, it is not feasible to visit each state and compute the value function directly. Therefore, we need to approximate the $Q$ and $V$ functions using function approximation techniques, such as deep neural networks. Despite the continuous nature of the state and action spaces, the Bellman equation can still be used to derive the optimal policy. This is known as the continuous Bellman equation, and it has been shown to be useful in finance applications (see Lemma \ref{continuous Bellman equation}).
\begin{lemma}\label{continuous Bellman equation}
(Continuous Bellman Equation) For a fixed policy $\pi$, we have the following recursive relation for the value function:
\begin{equation}
V^{\pi}(s)=\int_{\alpha \in A(s)} \pi(a|s) \int_{s^{\prime} \in S} \mathbb{P}(s'|s,a) \left[\boldsymbol{r}(\boldsymbol{s}, \boldsymbol{a},\boldsymbol{s}^{\prime})+\gamma V^{\pi}\left(s^{\prime}\right)\right] \text{d} \boldsymbol{s}^{\prime} \text{d} \boldsymbol{a}
\end{equation}
\end{lemma}

Another property of MDP and reinforcement learning is the Bellman optimality equation \citep{sutton1998introduction}. This proposition proves the existence of the optimal policy and the optimal value function. Formally, the Bellman optimality equation can be written as follows.
\begin{proposition}
(Bellman Optimality Equation) \citep{sutton1998introduction} There exists an optimal policy $\pi^{\star}$,
allowing the maximization of $V$-type and $Q$-type value functions at the same time. Formally written, we have that:
\begin{equation}
\pi^{\star}=\arg \max _{\pi \in \Pi} V^{\pi}(s)=\arg \max _{\pi \in \Pi} Q^{\pi}(s, a).
\end{equation}
\end{proposition}

\proof{Proof:}
To briefly show that there exists an optimal policy $\pi^{\star}$ that maximizes both the state value function $V^{\pi}(s)$ and the state-action value function $Q^{\pi}(s,a)$, we first define the optimal value functions $V^{\star}(s)$ and $Q^{\star}(s,a)$ as:
\begin{align*}
V^{\star}(\boldsymbol{s}) &= \max_{\pi \in \Pi} V^{\pi}(\boldsymbol{s}) \\
Q^{\star}(\boldsymbol{s}, \boldsymbol{a}) &= \max_{\pi \in \Pi} Q^{\pi}(\boldsymbol{s}, \boldsymbol{a}),
\end{align*}
where $\Pi$ is the set of all policies.

Now, we want to show that there exists a policy $\pi^{\star}$ that achieves the optimal value functions $V^{\star}(\boldsymbol{s})$ and $Q^{\star}(\boldsymbol{s},\boldsymbol{a})$ simutaneously.
We first prove that if $\pi'$ is a policy that achieves $V^{\star}(s)$, then it also achieves $Q^{\star}(\boldsymbol{s},\boldsymbol{a})$ for all $a \in \mathcal{A}(\boldsymbol{s})$, where $\mathcal{A}(\boldsymbol{s})$ is the set of actions available in state $\boldsymbol{s}$.

Assume that $\pi'$ achieves $V^{\star}(s)$, i.e., $V^{\pi'}(s) = V^{\star}(s)$. Then, we have:
\begin{equation}
\begin{aligned}
Q^{\pi'}(\boldsymbol{s}, \boldsymbol{a}) &= \sum_{\boldsymbol{s}' \in \mathcal{S}} \mathbb{P}(\boldsymbol{s}'| \boldsymbol{s},\boldsymbol{a}) [r(\boldsymbol{s}, \boldsymbol{a}, \boldsymbol{s}') + \gamma V^{\pi'}(\boldsymbol{s}')] \\
&\leq \sum_{s' \in S} \mathbb{P}(\boldsymbol{s}'| \boldsymbol{s},\boldsymbol{a})  [r(\boldsymbol{s}, \boldsymbol{a}, \boldsymbol{s}') + \gamma V^{\star}(\boldsymbol{s}')] \\
&= \max_{\pi \in \Pi} \sum_{s' \in S} \mathbb{P}(\boldsymbol{s}'| \boldsymbol{s},\boldsymbol{a})  [r(\boldsymbol{s}, \boldsymbol{a}, \boldsymbol{s}') + \gamma V^{\pi}(\boldsymbol{s}')] \\
&= Q^{\star}(\boldsymbol{s}, \boldsymbol{a})
\end{aligned}
\end{equation}
where we used the fact that $\pi'$ achieves $V^{\star}(\boldsymbol{s})$, and hence $V^{\pi'}(\boldsymbol{s}) \leq V^{\star}(\boldsymbol{s})$, and the maximization over policy space $\Pi$.
Similarly, we can show that if $\pi'$ is a policy that achieves $Q^{\star}(\boldsymbol{s},\boldsymbol{a})$, then it also achieves $V^{\star}(\boldsymbol{s})$. 
\Halmos
\endproof

This equation also indicates the following relations under optimal policy $\pi^{\star}$:
\begin{equation}\label{eq:Bellman optimality}
\begin{aligned}
&V^{\pi^{\star}}(\boldsymbol{s})=\max _{\boldsymbol{a} \in \mathcal{A}(\boldsymbol{s})}\left( \sum_{\boldsymbol{s}^{\prime}\in\mathcal{S}}\mathbb{P}(\boldsymbol{s}'| \boldsymbol{s},\boldsymbol{a}) r(\boldsymbol{s}, \boldsymbol{a},\boldsymbol{s}')+\gamma \sum_{\boldsymbol{s}^{\prime} \in \mathcal{S}} \mathbb{P}(\boldsymbol{s}'| \boldsymbol{s},\boldsymbol{a}) V^{\pi^{\star}}\left(\boldsymbol{s}'\right)\right) \\
&Q^{\pi^{\star}}(\boldsymbol{s}, \boldsymbol{a})=\sum_{\boldsymbol{s}^{\prime}\in\mathcal{S}}\mathbb{P}(\boldsymbol{s}'| \boldsymbol{s},\boldsymbol{a}) r(\boldsymbol{s}, \boldsymbol{a},\boldsymbol{s}')+\gamma \sum_{\boldsymbol{s}^{\prime} \in \mathcal{S}} \mathbb{P}(\boldsymbol{s}'| \boldsymbol{s},\boldsymbol{a}) \max _{\boldsymbol{a}^{\prime} \in \mathcal{A}(\boldsymbol{s}')} Q^{\pi^{\star}}\left(\boldsymbol{s}^{\prime}, \boldsymbol{a}^{\prime}\right) .
\end{aligned}
\end{equation}

Bellman optimality equation shows the existence of the optimal strategy. It also provides us with a way to find the optimal $\pi^{\star}$ by finding the optimal value functions $Q^{\pi^{\star}}(\boldsymbol{s}, \boldsymbol{a})$ or $V^{\pi^{\star}}(\boldsymbol{s})$. Since the aim of RL is to find the optimal policy $\pi^\star$ achieving the optimal value function, we can directly approximate the optimal value function bypassing modeling the transition $\mathbb{P}$.
The intuition of modeling the value function directly, rather than attempting to estimate the transition dynamics $\mathbb{P}(\boldsymbol{s}'|\boldsymbol{s},\boldsymbol{a})$ forms the basis of  \textit{model-free} RL algorithms \citep{silver2014ddpg, silver2016mastering, fujimoto2018td3}. These algorithms use various function types and techniques to approximate the value function induced by a given policy with the aim to achieve the equilibrium in Equaition \eqref{eq:Bellman optimality}. Our framework also employs model-free RL algorithms as the agents, given the difficulty of directly modeling the transition dynamics in financial markets. In the next section we will present the Twin Delayed Deep Deterministic Policy Gradient (TD3), which use neural networks to approximate the value function and policy.
 
\subsection{Twin Delayed Deep Deterministic Policy Gradient (TD3)}\label{ec:td3}
In this part we present the exact steps in deploying TD3 algorithm on the embedded states and the approaches to training a TD3 agent. The pseudo-code for TD3 algorithm based on the dynamic embedding is presented in Algorithm \ref{alg: TD3}.

As mentioned in the main body, the TD3 algorithm deploys two critic networks $Q_{\nu_1},Q_{\nu_2}: \mathcal{Z}\times\mathcal{A}\rightarrow\mathbb{R}$ to estimate the state-action value function, and use an actor network $\pi_\iota: \mathcal{S}\rightarrow\mathcal{A}$ to generate the best action. Each step we take action according to $\pi_\iota(\boldsymbol{z_s})$, and evaluate the action based on the value function $Q_{\nu}$. The target networks $Q_{\nu'_1},Q_{\nu'_2},\pi_\iota'$ are used for stable updates of the parameters of the original network, which we will discuss later.

The Twin Delayed Deep Deterministic policy gradient (TD3) algorithm builds upon the Deep Deterministic Policy Gradient (DDPG) algorithm \citep{lillicrap2015continuous}. TD3 addresses issues such as overestimation bias and learning instability by incorporating improvements in the form of \textit{double Q-learning}, \textit{delayed policy updates}, and \textit{target policy smoothing}.

\subsubsection*{Double Q-learning.}
In the TD3 algorithm, double Q-learning is employed to mitigate the overestimation bias that may arise from using a single critic network. Overestimation can lead to suboptimal policies and slower convergence. TD3 incorporates two separate critic networks, $Q_{\nu_1}$ and $Q_{\nu_2}$, to independently learn the Q-values. During the update process, the algorithm takes the minimum of the two critics' predictions ($\min_{i}Q_{\nu_i}$), which reduces overestimation bias. This technique is inspired by Double Q-learning, and it helps improve the stability of the learning process.

\subsubsection*{Delayed Policy Updates.}
TD3 introduces delayed policy updates to further enhance the stability of training. In this approach, the actor network is updated less frequently compared to the critic networks. As shown in the steps :\textit{Update actor network} and \textit{Smoothly update three target network} in Algorithm \ref{alg: TD3}, we only update the actor network every $d$ step. By delaying the policy updates, the actor has access to more accurate Q-value estimates, which results from the additional critic network updates. This, in turn, leads to more stable policy improvements and prevents premature convergence to suboptimal policies.

\subsubsection*{Target Policy Smoothing.}
Target policy smoothing is a technique introduced in TD3 to address the exploitation of unrealistic Q-value estimates due to the inherent function approximation errors in neural networks. This method involves adding a small amount of noise $\varepsilon$ (in Step 2 in Algorithm \ref{alg: TD3}) to the target action during the critic update step, effectively smoothing the target policy. By incorporating noise into the target actions, the algorithm learns to generalize better across different states and actions, reducing the chance of overfitting to specific scenarios.

Together, these three improvements enable the TD3 algorithm to exhibit superior performance and stability compared to the DDPG algorithm. The incorporation of double Q-learning, delayed policy updates, and target policy smoothing helps to mitigate overestimation bias and learning instability, leading to more reliable and efficient reinforcement learning.

Throughout the training process, TD3 alternates between learning the actor and critic networks using experience replay and soft updates to the target networks. This iterative procedure allows TD3 to achieve superior performance and stability compared to its predecessor, DDPG.
\begin{algorithm}[htbp]
    \caption{TD3 algorithm with dynamic embedding}
    \label{alg: TD3}
    \SetAlgoLined
\KwIn{Initialize critic networks $Q_{\nu_{1}}, Q_{\nu_{2}}$, and actor network $\pi_{\iota}$ with parameters $\nu_{1}, \nu_{2}, \iota$; initialize target networks $\nu_{1}^{\prime} \leftarrow \nu_{1}, \nu_{2}^{\prime} \leftarrow \nu_{2}, \iota^{\prime} \leftarrow \iota$; initialize buffer $\mathcal{B}$; set action noise variance $\sigma^2$ and maximum noise level $c$; initialize the pre-trained encoder $\Gamma_\phi$; set frequency of delayed policy update $d$ and frequency of dynamic update |U|}
    \For{$t=1$ to $T$}{
\textbf{Take action}: receive states $\boldsymbol{s}$, encode it with $\Gamma_\phi$ into $\boldsymbol{z_s}$, select action with exploration noise $\boldsymbol{a} \sim \pi_{\phi}(\boldsymbol{z_s})+\epsilon$, $\epsilon \sim \mathcal{N}(0, \sigma^2I)$, observe reward $r$ and new state $\boldsymbol{s^{\prime}}$; and store transition tuple $\left(\boldsymbol{s}, \boldsymbol{a}, r, \boldsymbol{s^{\prime}}\right)$ in $\mathcal{B}$;\\
\textbf{Update critic networks}: sample mini-batch of $b$ transitions $\{\left(\boldsymbol{s}_i, \boldsymbol{a}_i, r_i, \boldsymbol{s^{\prime}}_i\right)\}_{i=1}^{b}$ from $\mathcal{B}$ and encode the corresponding latent state $\left\{\boldsymbol{z_{s_i}}\right\}_{i=1}^b$ with $\Gamma_\phi$;\\
    recalculate the target optimal action under the current policy network
    $$\tilde{\boldsymbol{a}}_i \leftarrow \pi_{\iota^{\prime}}\left(\boldsymbol{z_{s_i}}(\boldsymbol{s^{\prime}})\right)+\epsilon, \quad \epsilon \sim \operatorname{clip}(\mathcal{N}(0, \sigma^2I),-c\boldsymbol{1}, c\boldsymbol{1}),$$ and the target value:
    $y_i \leftarrow r_i +\gamma \min _{i=1,2} Q_{\nu_{i}^{\prime}}\left(\boldsymbol{s^{\prime}}, \tilde{\boldsymbol{a}}\right)$; and update parameter of critic networks: $$\nu_{i} \leftarrow \operatorname{argmin}_{\nu_{i}} b^{-1} \sum\left(y_i-Q_{\nu_{i}}(\boldsymbol{z_{s_i}}, \boldsymbol{a})\right)^{2}.$$ 
\If {$t = 0 \  (\bmod \ d) $}{
\textbf{Update actor network} parameter through:
$$\iota =\iota-\alpha_\iota \left.b^{-1} \sum \nabla_{\boldsymbol{a}} Q_{\nu_{1}}(\boldsymbol{z_s}, \boldsymbol{a})\right|_{\boldsymbol{a}=\pi_{\iota}(\boldsymbol{z_s})} \nabla_{\iota} \pi_{\iota}(\boldsymbol{z_s}).$$
\textbf{Smoothly update target network} parameters: 
\begin{align*}
   \nu_{i}^{\prime} & \leftarrow \tau \nu_{i}+(1-\tau) \nu_{i}^{\prime}, \\
   \iota^{\prime} & \leftarrow \tau \iota+(1-\tau) \iota^{\prime}.
\end{align*}
}
\If{$t = 0 \ (\bmod \ |U|)$}{\textbf{Update the encoder} $\Gamma_\phi$ with online meta-learning.}}
  \end{algorithm}

\section{Details on Generative Autoencoders}\label{ec:autoencoders}
Generative autoencoders aim to simulate the real-world data generation process of a random variable by learning the joint distribution of all variables related to its generation. 
% We can tackle the generation process through the point of optimal transport according to \cite{bousquet2017optimal} and \cite{Martin2017WGAN}.

\subsection{Generative Autoencoders}
We begin by introducing the concept and use of generative encoders. Considering two random variables $X$ and $Y$ in a  compact set $
\mathcal{X}$, we aim to learn the joint distribution of two random variables $X\in\mathcal{X}$ and $Y\in\mathcal{X}$ through a latent variable $Z\in \mathcal{Z}$ that supports the conditional independence of $X$ and $Y$, such that $(Y \bot X) \mid Z$. Most work using generative models focuses on regenerating $X\in \mathcal{X}$ through the latent variable $Z\in\mathcal{Z}$, or what is known as \textit{self-regeneration} ($X\xrightarrow{\text{encoder}} Z \xrightarrow{\text{decoder}}X$). The step $X\xrightarrow{\text{encoder}} Z$ is usually called encoding and $Z\xrightarrow{encoder} X$ is usually called decoding. 

We describe the generative encoders with a probabilistic model. For $X,Y \in \mathcal{X}$ and $Z\in\mathcal{Z}$, we assume $X \sim P_{X}$ and $Y \sim P_{Y}$, and for $Z\in P_{Z}$.
We use $\mathcal{P}_{X,Y}:=\mathcal{P}\left(X \sim P_{X}, Y \sim P_{Y}\right)$ to denote the set of all joint distributions of $(X, Y)$ with marginal distributions $P_{X}$ and $P_{Y}$. 
%The subscript $G$ in $P_{G}$ represents the generative model, which is typically represented by a deep neural network. 
Similarly, for a pair $(X,Z)$, we have for joint distribution $\mathcal{P}_{X,Z}:=\mathcal{P}\left(X \sim P_{X}, Z \sim P_{Z}\right)$, where $P_{Z}$ is the prior distribution over the latent variable $\mathcal{Z}$. 
In the context of generative tasks, we define a cost function $\mathcal{C}(X, Y): \mathcal{X} \times \mathcal{X} \rightarrow \mathcal{R}_{+}$, which measures the difference or regeneration loss between the input data $X$ and the generated data $Y$. The objective is to minimize the transport cost
\begin{equation} \label{equation: transportation loss}
W_{\mathcal{C}}(P_X, P_Y):=\inf _{P_{X,Y} \in \mathcal{P}_{X,Y}}\mathbb{E}_{(X, Y) \sim P_{X,Y}} [\mathcal{C}(X, Y)].
\end{equation}

%The set of all $(X, Y, Z)$ joint distributions is denoted by $\mathcal{P}_{X, Y, Z}$, with $X \sim P_{X},(Y, Z) \sim P_{Y, Z}$, and $(Y \bot X) \mid Z$. Additionally, we use the notation $\mathcal{P}_{X, Y}$ and $\mathcal{P}_{X, Z}$ to denote the sets of marginal distributions on $(X, Y)$ and $(X, Z)$ induced by the distribution $\mathcal{P}_{X, Y, Z}$. 

Our setting, as mentioned in the main body of the paper, is modified into
\begin{equation}\label{eq:ec transition equation}
\boldsymbol{s} \xrightarrow[\text{encode}]{\Gamma_\phi} \boldsymbol{\boldsymbol{z_s}} \xrightarrow[\text{\text{decode with }} \boldsymbol{a}\in\mathcal{A}(s)]{G_\theta} \boldsymbol{s^\prime},
\end{equation}
where each notation represents:
\begin{itemize}
\item Encoder $\Gamma_\phi(\boldsymbol{z_s}|\boldsymbol{s})$: maps the state $\boldsymbol{s}$ to its latent representation $z_s$,  parameterized by $\phi$;
\item Decoder $G_\theta(\boldsymbol{s}'|\boldsymbol{z_s}, \boldsymbol{a})$: maps the  latent state $\boldsymbol{z_s}$ and action $\boldsymbol{a}$ back to the predicted next state $\boldsymbol{\hat{s}}'$, parameterized by $\theta$.
\end{itemize}
With the encoder and decoder, we reconstruct the next state through
\begin{equation}
    \mathbb{P}'_{\theta,\phi}(\boldsymbol{\hat{s}}|\boldsymbol{s},\boldsymbol{a})= \int_{z_s\in\mathcal{Z}} G_\theta(\boldsymbol{s}'|\boldsymbol{z_s}, \boldsymbol{a})\Gamma_\phi(\boldsymbol{z_s}|s)\text{d}\boldsymbol{z_s}.
\end{equation}
Similar to conventional autoencoders, we hope that for any given $(\boldsymbol{s},\boldsymbol{a})$, the reconstructed distribution $\mathbb{P}'_{\theta,\phi}$ of next state $\boldsymbol{s}$ is close to the real distribution $\mathbb{P}_{\theta,\phi}$.
Thus we modify \eqref{equation: transportation loss} to minimize the transport cost between the reconstructed transition probability and the exact transition probability under given $(\boldsymbol{s},\boldsymbol{a})$. We use $\mathcal{P}_{\boldsymbol{\hat{s}},\boldsymbol{s}}:= \mathcal{P}(\boldsymbol{s}'\sim \mathbb{P}(\cdot|\boldsymbol{s},\boldsymbol{a}),\boldsymbol{\hat{s}}'\sim \mathbb{P}'_{\theta,\phi}(\cdot|\boldsymbol{s},\boldsymbol{a}))$ denote all distribution of $(\boldsymbol{\hat{s}}',\boldsymbol{s}')$ with marginal distribution of $\mathbb{P}$ and $\mathbb{P}'_{\theta,\phi}$, thus we have:
\begin{equation} \label{equation: transition transportation loss}
W_{\mathcal{C}}(\mathbb{P}'_{\theta,\phi}, \mathbb{P}):=\inf _{P_{\boldsymbol{\hat{s}},\boldsymbol{s}} \in \mathcal{P}_{\boldsymbol{\hat{s}},\boldsymbol{s}}}\mathbb{E}_{\boldsymbol{\hat{s}},\boldsymbol{s} \sim P_{\boldsymbol{\hat{s}},\boldsymbol{s}}} [\mathcal{C}(\boldsymbol{\hat{s}},\boldsymbol{s})].
\end{equation}
In our framework, the decoder $G_\theta(\cdot|\boldsymbol{z_s},\boldsymbol{a})$ for next state is constructed through two steps, where the $\boldsymbol{z_s}$ is sampled from a fixed distribution $P_{\boldsymbol{z}}$ on the latent space $\mathcal{Z}$, and then $\boldsymbol{z_s}$ is mapped to the next state $\hat{\boldsymbol{s}'}\in\mathcal{S}$, leading to 
\begin{equation}
\mathbb{P}_\theta'(\boldsymbol{s}'|\boldsymbol{s},\boldsymbol{a})=\int_{\mathcal{Z}} G_\theta(\boldsymbol{s}'|\boldsymbol{z_s},\boldsymbol{a})p_z(\boldsymbol{z_s})\text{d}\boldsymbol{z_s}.
\end{equation}
There is a special case that the decoder is a deterministic decoder, which indicates the $G_\theta(\cdot|\boldsymbol{s},\boldsymbol{a})$ is Dirac distribution. For simplicity, in deterministic decoder case, we write the decoder as $\boldsymbol{s}'=f_{G_\theta}(\boldsymbol{s},\boldsymbol{a})$, which deterministically maps the state-action pair $(\boldsymbol{s},\boldsymbol{a})$ to the next state $\boldsymbol{s}'$.

Then we have the marginal distribution of $\boldsymbol{z_s}$ under given $\boldsymbol{s}$ from the state encoder $\mathbb{P}_\phi(\boldsymbol{z_s})=\Gamma_\phi(\boldsymbol{z_s}|\boldsymbol{s})$, which is assumed to match the known prior $P_z$ for any given $\boldsymbol{s}$.
\begin{proposition}\label{prop:encode-decode}
With given current state-action pair $(\boldsymbol{s},\boldsymbol{a})$, for $G_\theta$ as defined above with deterministic $G_\theta(\boldsymbol{s}'\mid\boldsymbol{z_s},\boldsymbol{a})$ and any function $f_{G_\theta}(\boldsymbol{z_s},\boldsymbol{a}): \mathcal{Z}\times\mathcal{A} \rightarrow \mathcal{S}$, we sample $\boldsymbol{z_s}$ from $\Gamma_\phi(\boldsymbol{z_s}|\boldsymbol{s})$ and then  
$\boldsymbol{\hat{s}}'$ from $G_\theta(\boldsymbol{z_s},\boldsymbol{a})$. Comparing the estimated next state $\boldsymbol{\hat{s}}'$ and the exact next state $\boldsymbol{s}'$,
we have 
\begin{equation}
\inf _{P \in \mathcal{P}\left(\boldsymbol{s}' \sim \mathbb{P}, \boldsymbol{\hat{s}}'\sim \mathbb{P}'_{\phi,\theta}\right)} \mathbb{E}_{(\boldsymbol{s}', \boldsymbol{\hat{s}}') \sim P}[\mathcal{C}(\boldsymbol{s}', \boldsymbol{\hat{s}}')]=\inf _{\Gamma_\phi(\cdot|\boldsymbol{s})=P_Z} \mathbb{E}_{\mathbb{P}_{\boldsymbol{s}'}} \mathbb{E}_{\Gamma_\phi(\boldsymbol{z_s} \mid \boldsymbol{s})}[\mathcal{C}(\boldsymbol{s}', f_{G_\theta}(\boldsymbol{z_s},\boldsymbol{a}))].
\end{equation}
\end{proposition}

\textit{Proof:}  
Since the $G_\theta(\boldsymbol{\hat{s}}'|\boldsymbol{z_s},\boldsymbol{a})$ is a deterministic mapping, for any subset $S$ in $\mathcal{S}$, we have $\mathbb{E}\left[\mathbb{I}_{[\boldsymbol{\hat{s}}'\in S]}|\boldsymbol{s}',\boldsymbol{a},\boldsymbol{z_s}\right] = \mathbb{E}\left[\mathbb{I}_{[\boldsymbol{\hat{s}}'\in S]}|\boldsymbol{a},\boldsymbol{z_s}\right]$, which implies the independence of 
$\boldsymbol{\hat{s}}'$ and $ \boldsymbol{s}'$ given $(\boldsymbol{z_s},\boldsymbol{a})$. 

The tower rule of expectation, and the conditional independence property of $\mathcal{P}_{\boldsymbol{s}', \boldsymbol{\hat{s}}', \boldsymbol{z_s}}$ with given $(\boldsymbol{s},\boldsymbol{a})$, we have that 
\begin{equation}
\begin{aligned}
W_\mathcal{C}\left(\mathbb{P}'_{\theta,\phi}, \mathbb{P}\right) & =\inf _{P \in \mathcal{P}_{\boldsymbol{s}',\boldsymbol{z_s},\boldsymbol{\hat{s}}'}}\mathbb{E}_{P(\boldsymbol{s}',\boldsymbol{z_s},\boldsymbol{\hat{s}}'|\boldsymbol{s},\boldsymbol{a}) }[\mathcal{C}(\boldsymbol{s}', \boldsymbol{\hat{s}}')] \\
& =\inf _{P \in \mathcal{P}_{\boldsymbol{s}',\boldsymbol{z_s},\boldsymbol{\hat{s}}'}} \mathbb{E}_{\Gamma_\phi{(\boldsymbol{z_s}|\boldsymbol{s})}} \mathbb{E}_{\boldsymbol{s}' \sim P(\boldsymbol{s}' \mid \boldsymbol{s},\boldsymbol{a})} \mathbb{E}_{\boldsymbol{\hat{s}}' \sim G_\theta(\boldsymbol{\hat{s}}'  \mid \boldsymbol{z_s},\boldsymbol{a})}[\mathcal{C}(\boldsymbol{s}, \boldsymbol{\hat{s}}')] \\
& =\inf _{P \in \mathcal{P}_{\boldsymbol{s}',\boldsymbol{z_s},\boldsymbol{\hat{s}}'}} \mathbb{E}_{\Gamma_\phi{(\boldsymbol{z_s}|\boldsymbol{s})}} \mathbb{E}_{\boldsymbol{s}' \sim P(\boldsymbol{s}' \mid \boldsymbol{s},\boldsymbol{a})}[\mathcal{C}(\boldsymbol{s}', f_{G_\theta}(\boldsymbol{z_s},\boldsymbol{a}))] \\
& =\inf _{P_{\boldsymbol{s}', \boldsymbol{z_s}} \in \mathcal{P}_{\boldsymbol{s}', \boldsymbol{z_s}}} \mathbb{E}_{(\boldsymbol{s}', \boldsymbol{z_s}) \sim P_{\boldsymbol{s}', \boldsymbol{z_s}}}[\mathcal{C}(\boldsymbol{s}', f_{G_\theta}(\boldsymbol{z_s},\boldsymbol{a}))] \\
& = \inf_{\Gamma_\phi(\cdot|\boldsymbol{s})=P_Z}  \mathbb{E}_{\mathbb{P}_{\boldsymbol{s}'}} \mathbb{E}_{\Gamma_\phi(\boldsymbol{z_s} \mid \boldsymbol{s})}[\mathcal{C}(\boldsymbol{s}', f_{G_\theta}(\boldsymbol{z_s},\boldsymbol{a}))].
\end{aligned}
\end{equation}
It remains to notice that $\mathcal{P}_{\boldsymbol{s}', \boldsymbol{z_s}}=\mathcal{P}\left(\boldsymbol{s}' \sim \mathbb{P}(\cdot|\boldsymbol{s},\boldsymbol{a}), \boldsymbol{z_s} \sim P_Z\right)$.
\begin{corollary}\label{coro: random decoder}
With given current state-action pair $(\boldsymbol{s},\boldsymbol{a})$, for $G_\theta$ as defined above with random decoder $G_\theta(\boldsymbol{s}'\mid\boldsymbol{z_s},\boldsymbol{a})$, if the mean value function $f_{G_\theta}(\boldsymbol{z_s},\boldsymbol{a}): \mathcal{Z}\times\mathcal{A} \rightarrow \mathcal{S}$, we sample $\boldsymbol{z_s}$ from $\Gamma_\phi(\boldsymbol{z_s}|\boldsymbol{s})$ and then  
$\boldsymbol{\hat{s}}'$ from $G_\theta(\boldsymbol{z_s},\boldsymbol{a})$. If adopting the squared loss distance $\mathcal{C}(\boldsymbol{s}',\boldsymbol{\hat{s}}')=\|\boldsymbol{s}'-\boldsymbol{\hat{s}}'\|_2^2$ and assuming the variance associated with prior $P_Z$ as $\{\sigma_i^2\}_{i=1}^{\dim(\mathcal{Z})}$, we have that
\begin{equation}
\inf _{P \in \mathcal{P}\left(\boldsymbol{s}' \sim \mathbb{P}, \boldsymbol{\hat{s}}'\sim \mathbb{P}'_{\phi,\theta}\right)} \mathbb{E}_{(\boldsymbol{s}', \boldsymbol{\hat{s}}') \sim P}[\mathcal{C}(\boldsymbol{s}', \boldsymbol{\hat{s}}')]=
\sum_{i=1}^{\dim(\mathcal{Z})}\sigma_i^2+ \inf _{\Gamma_\phi(\cdot|\boldsymbol{s})=P_Z} \mathbb{E}_{\mathbb{P}_{\boldsymbol{s}'}} \mathbb{E}_{\Gamma_\phi(\boldsymbol{z_s} \mid \boldsymbol{s})}[\|\boldsymbol{s}'-f_{G_\theta}(\boldsymbol{z_s},a)\|_2^2].
\end{equation}
\end{corollary}

\textit{Proof:} With squared distance, we can rewrite the \ref{equation: transition transportation loss} as
\begin{equation}
\begin{aligned}
W_\mathcal{C}\left(\mathbb{P}'_{\theta,\phi}, \mathbb{P}\right) & =\inf _{P \in \mathcal{P}_{\boldsymbol{s}',\boldsymbol{z_s},\boldsymbol{\hat{s}}'}}\mathbb{E}_{P(\boldsymbol{s}',\boldsymbol{z_s},\boldsymbol{\hat{s}}'|\boldsymbol{s},\boldsymbol{a}) }[\|\boldsymbol{s}'-\boldsymbol{\hat{s}}'\|_2^2] \\
&= \inf _{P \in \mathcal{P}_{\boldsymbol{s}',\boldsymbol{z_s},\boldsymbol{\hat{s}}'}}\mathbb{E}_{P(\boldsymbol{s}',\boldsymbol{z_s},\boldsymbol{\hat{s}}'|\boldsymbol{s},\boldsymbol{a}) }[\|\boldsymbol{s}'-f_{G_\theta}(\boldsymbol{a},\boldsymbol{z_s})+f_{G_\theta}(\boldsymbol{a},\boldsymbol{z_s})-\boldsymbol{\hat{s}}'\|_2^2]\\
& = \inf _{P \in \mathcal{P}_{\boldsymbol{s}',\boldsymbol{z_s},\boldsymbol{\hat{s}}'}}\mathbb{E}_{P(\boldsymbol{s}',\boldsymbol{z_s},\boldsymbol{\hat{s}}'|\boldsymbol{s},\boldsymbol{a}) }[\|\boldsymbol{s}'-f_{G_\theta}(\boldsymbol{a},\boldsymbol{z_s})\|_2^2+2\left<\boldsymbol{s}'-f_{G_\theta},f_{G_\theta}-\boldsymbol{\hat{s}}'\right>]+\sum_{i=1}^{\dim(\mathcal{Z})}\sigma_i^2\\
& = \sum_{i=1}^{\dim(\mathcal{Z})}\sigma_i^2+ \inf _{\Gamma_\phi(\cdot|\boldsymbol{s})=P_Z} \mathbb{E}_{\mathbb{P}_{\boldsymbol{s}'}} \mathbb{E}_{\Gamma_\phi(\boldsymbol{z_s} \mid \boldsymbol{s})}[\|\boldsymbol{s}'-f_{G_\theta}(\boldsymbol{z_s},a)\|_2^2],
\end{aligned}
\end{equation}
which concludes the proof.
\Halmos
\endproof

Proposition \ref{prop:encode-decode} and Corollary \ref{coro: random decoder} allow us to find the optimal encoder-decoder if the encoder $\Gamma_\phi(\cdot|\boldsymbol{s})$ can map the raw states into a known prior $P_Z$. In order to find a numerical solution, we can relax the constraint $\Gamma_\phi = P_Z$ by adding a penalty term and minimize the loss function:
\begin{equation}\label{eq:relaxed WAE}
\inf_{\theta,\phi} \mathcal{L}_{\theta,\phi}= \inf_{\Gamma_\phi(\cdot|\boldsymbol{s}),f_{G_\theta}}  \mathbb{E}_{\mathbb{P}_{\boldsymbol{s}'}} \mathbb{E}_{\Gamma_\phi(\boldsymbol{z_s} \mid \boldsymbol{s})}[\mathcal{C}(\boldsymbol{s}', f_{G_\theta}(\boldsymbol{z_s},\boldsymbol{a}))] + \lambda \text{Dist} (\Gamma_\phi(\cdot|\boldsymbol{s}),P_Z),
\end{equation}
where $\text{Dist} (\Gamma_\phi(\cdot|\boldsymbol{s}),P_Z)$ measure the divergence between distribution $\Gamma_\phi(\cdot|\boldsymbol{s})$ and $P_Z$. There are various types of divergence metrics, such as maximum mean discrepancy (MMD) mentioned in the main body and generative-adversarial-network (GAN-based) distance. With different divergence metrics and some modifications, we obtain the corresponding autoencoders. The following are some commonly used autoencoders.
\subsection{Different Generative Models and Loss Functions}
Well-known generative models include generative adversarial network (GAN), Wasserstein GAN, variational autoencoder, and Wasserstein autoencoder. The encoder parts in all of them can be fitted in our framework as the embedding.
% Using the techniques in \S \ref{sec: optimization approach}, we can derive the loss functions of these auto-encoders. 

\textbf{Wasserstein autoencoder (WAE)} \citep{tolstikhin2018wasserstein} WAE applies the direct adaption on Equation \eqref{eq:relaxed WAE}, which optimizes on the loss function with relaxed the constraints on $\Gamma_\theta(\boldsymbol{z_s}|\boldsymbol{s})$ by adding a penalty to the objective:
\begin{equation}
\mathcal{L}_{\mathrm{WAE}}(\theta,\phi)= \mathbb{E}_{\mathbb{P}_{\boldsymbol{s}'}} \mathbb{E}_{\Gamma_\phi(\boldsymbol{z_s} \mid \boldsymbol{s})}[\mathcal{C}(\boldsymbol{s}', f_{G_\theta}(\boldsymbol{z_s},\boldsymbol{a}))] + \lambda \text{Dist} (\Gamma_\phi(\cdot|\boldsymbol{s}),P_Z).
\end{equation}
Considering using the MMD as the discrepancy metric, we provide a numerical algorithm to minimize the loss function. The algorithm is presented in \ref{alg:WAE}.

\RestyleAlgo{ruled}
\begin{algorithm}[htbp]
  \caption{Training WAE}
  \label{alg:WAE}
  \SetAlgoLined
  \KwIn{Regularization coefficient $\lambda>0$, characteristic kernel $k$, prior $P_Z$, training set $\mathcal{B}$.}
  Initialize encoder $\Gamma_{\phi}$, decoder $G_{\theta}$\;
  \While{$(\phi, \theta)$ not converged}{
    Sample batch $\{(\boldsymbol{s},\boldsymbol{a},\boldsymbol{s^\prime})_i\}_{i=1}^n$ from $\mathcal{B}$\;
    Sample $\{\boldsymbol{z}_1, \ldots, \boldsymbol{z}_n\}$ from $P_Z$\;
    Sample $\{\boldsymbol{\tilde{z}}_i\}_{i=1}^n$ where $\boldsymbol{\tilde{z}}_i \sim \Gamma_{\phi}(\cdot|\boldsymbol{s}_i)$\;
    Compute loss:
    \begin{equation}
    \begin{aligned}
    \mathcal{L} = \frac{1}{n} \sum_{i=1}^n c\left(\boldsymbol{s^\prime_i}, G_{\theta}(\boldsymbol{\tilde{z}_i}, \boldsymbol{a_i})\right) 
    + \frac{\lambda}{n(n-1)} \sum_{\ell \neq j} \left[k(\boldsymbol{z_\ell}, \boldsymbol{z_j}) + k(\boldsymbol{\tilde{z}_\ell}, \boldsymbol{\tilde{z}_j})\right] 
    - \frac{2\lambda}{n^2} \sum_{\ell, j} k(\boldsymbol{z_\ell}, \boldsymbol{\tilde{z}_j})
    \end{aligned}
    \end{equation}
    Update $\Gamma_{\phi}$ and $G_{\theta}$ by backpropagating $\mathcal{L}$\;
  }
\end{algorithm}

\textbf{Generative adversarial network (GAN).} \citep{goodfellow2014generative} With given $(\boldsymbol{s},\boldsymbol{a})$, GAN trys to minimizes the 
\begin{equation}
\mathcal{L}_{\mathrm{GAN}}\left(\phi,\theta\right)=\sup _{T \in \mathcal{T}} \mathbb{E}_{\boldsymbol{s}' \sim \mathbb{P}(\cdot|\boldsymbol{s},\boldsymbol{a})}[\log T(\boldsymbol{s}')]+\mathbb{E}_{z_s' \sim \Gamma_\phi(\boldsymbol{z_s}|\boldsymbol{s})}[\log (1-T(f_{G_\theta}(\boldsymbol{z_s},\boldsymbol{a})))]
\end{equation}
with respect to a deterministic generator $f_{G_\theta}(\boldsymbol{z_s},\boldsymbol{a}): \mathcal{Z}\times\mathcal{A} \rightarrow \mathcal{S}$, where $\mathcal{T}$ is any non-parametric class of choice.

\textbf{Wasserstein GAN (WGAN)} \citep{Martin2017WGAN} The WGAN minimizes 
\begin{equation}\label{WGAN}
\mathcal{L}_{\mathrm{WGAN}}\left(\phi,\theta\right)=\sup _{T \in \mathcal{W}} \mathbb{E}_{\boldsymbol{s}' \sim \mathbb{P}(\cdot|\boldsymbol{s},\boldsymbol{a})}[T(\boldsymbol{s}')]-\mathbb{E}_{z_s' \sim \Gamma_\phi(\boldsymbol{z_s}|\boldsymbol{s})}[ T(f_{G_\theta}(\boldsymbol{z_s},\boldsymbol{a}))],
\end{equation}
where $\mathcal{W}$ is any subset of $1$-Lipschitz functions on $\mathcal{S}$. 

\textbf{Variational autoencoder (VAE)} \citep{Kingma2014AutoEncodingVB} With given $(\boldsymbol{s},\boldsymbol{a})$ tries to minimize the loss $\mathcal{L}_{\mathrm{VAE}}$
\begin{equation}
\mathcal{L}_{\mathrm{VAE}}\left(\phi,\theta\right)=\inf _{\Gamma_\phi(\boldsymbol{z_s} \mid \boldsymbol{s}) \in \Gamma} D_{\mathrm{KL}}\left(\Gamma_\phi(\boldsymbol{z_s} \mid \boldsymbol{s}), P_{Z}\right)-\mathbb{E}_{\Gamma_\phi(\boldsymbol{z_s} \mid \boldsymbol{s})}\left[\log {G}_\theta(\boldsymbol{\hat{s}}' \mid \boldsymbol{z_s},\boldsymbol{a})\right]
\end{equation}
where the $\Gamma$ is a known family of distribution. Here we provide a sample approach to train VAE in Algorithm \ref{alg:VAE}.

\RestyleAlgo{ruled}
\begin{algorithm}[htbp]
    \caption{Training $\beta-$VAE (with $\beta=1$)}
    \label{alg:VAE}
    \SetAlgoLined
     \KwIn{Characteristic positive-definite kernel $k$.
Initialize the parameters of the encoder $\Gamma_{\phi}$, decoder $G_{\theta}$, collected dataset (memory buffer) $\mathcal{D}=\{(\boldsymbol{s},\boldsymbol{a},\boldsymbol{s^\prime})_i\}^N_{i=1}$.}
    \While{$(\phi, \theta)$ not converged}{Sample $\left\{(\boldsymbol{s},\boldsymbol{a},\boldsymbol{s^\prime})_1, \ldots, (\boldsymbol{s},\boldsymbol{a},\boldsymbol{s^\prime})_n\right\}$ from $\mathcal{D}$;\\
      \ForEach{$i=1, \ldots, n$}{
Sample $\boldsymbol{\tilde{z}}_{i}$ from $\Gamma_{\phi}\left(Z \mid \boldsymbol{s_{i}}\right)$; \\
Sample $\boldsymbol{\hat{s}'_{i}}$ from $G_{\theta}\left(\boldsymbol{\tilde{z}}_{i},\boldsymbol{a_i}\right)$; \\
Update $Q_{\phi}$ and $G_{\theta}$ by backward propagating the loss with SGD:
$$\frac{1}{n} \sum_{i=1}^{n} \left\|\boldsymbol{s^\prime_{i}}- \hat{\boldsymbol{s}}'\right\|^2_2+\frac{1}{2} \sum_{j=1}^{\dim(\mathcal{Z})}\left(1+\log \left(\left(\sigma_{j}\right)^{2}\right)-\left(\mu_{j}\right)^{2}-\left(\sigma_{j}\right)^{2}\right),$$
where the  $\sigma_j$ and $\mu_j$ represent the j-th element in $\boldsymbol{\mu}$ and $\boldsymbol{\sigma}$.}}
\end{algorithm}

\section{Addition to Experiments}
In this section, we present some additional information regarding our numerical experiments.
\subsection{Stock Characteristics} \label{ec:characteristics}
In this section, we provide a brief description of the eight characteristics that are used in this study. We calculate the characteristics using a rolling window approach. The sizes of the window, $n$, are selected as 7, 14, and 30 calendar days, with a minimum requirement of 3, 6, and 10 valid observations within the rolling window, respectively. Table \ref{tab: characteristics} outlines the methods for constructing these stock characteristics.

\begin{comment}

The \cite{amihud02} Illiquid measure is defined as 
\[
\mathrm{ILLQ}_t = \frac{1}{n}\sum_{i=0}^{n-1}\frac{
\left\vert R_{t-i}\right\vert }{\$V_{t-i}}\times 10^{6},
\]
where $R_t$ is the return, and $\$V_t$ is the dollar volume on day $t$.

The bid-ask spread can be calculated as 
\[
\mathrm{SPREAD}_t = \frac{1}{n}\sum_{i=0}^{n-1}\frac{\left(
askp_{t-i}-bidp_{t-i}\right) }{\left( askp_{t-i}+bidp_{t-i}\right) /2},
\]
where $askp_t$ and $bidp_t$ are closing ask and bid prices on day $t$ respectively.

The share turnover is given by 
\[
\mathrm{TURN}_t = \frac{1}{n}\sum_{i=0}^{n-1}\frac{V_{t-i}}{
\mathrm{Shrout}_{t-i}},
\]
where $V_t$ and $\mathrm{Shrout}_t$ are trading volume and shares outstanding on day $t$ respectively.

The number of days with zero trades over $n$ days is given by 
\[
\mathrm{ZTrade}_t =  \sum_{i=0}^{n-1}1_{\left\{ V_{t-i}=0\right\} }.
\]

The price trend during the past $n$ days can be calculated as 
\[
\mathrm{TREND}_t = \Pi_{i=0}^{n-1}(1+R_{t-i})-1
\]
\end{comment}

\begin{table}[htbp]
\TABLE
{Stock characteristics \label{tab: characteristics}}
{
\begin{tabular}{lll}
\toprule
Characteristics & Abbreviation & Formula \\
\midrule
\cite{amihud02} illiquidity & Illiq & $\frac{1}{n}\sum_{s=0}^{n-1}\frac{
\left\vert R_{t-s}\right\vert }{\$V_{t-s}}\times 10^{6}$ \\
Bid-ask spread  & Spread & $\frac{1}{n}\sum_{s=0}^{n-1}\frac{\left(
askp_{t-s}-bidp_{t-s}\right) }{\left( askp_{t-s}+bidp_{t-s}\right) /2}$ \\
Share turnover & Turn & $\frac{1}{n}\sum_{s=0}^{n-1}\frac{V_{t-s}}{
\mathrm{shrout}_{t-s}}$ \\
\#Days with zero trades & Ztrade & $\sum_{s=0}^{n-1}1_{\left\{ V_{t-s}=0\right\} }$ \\
Price Trend & Trend & $\Pi_{s=0}^{n-1}(1+R_{t-s})-1$ \\
Return Volatility & Retvol & $\sqrt{\frac{1}{n-1}\sum_{s=0}^{n-1}\left(
R_{t-s}-\mu _{t}\right) ^{2}}$ , where $\mu _{t}=\frac{1}{n}
\sum_{s=0}^{n-1}R_{t-s}$ \\
Market beta & Beta & Market coefficient of CAPM model \\
Idiosyncratic volatility &  Ivol & $\sqrt{\frac{1}{n-1}\sum_{s=0}^{n-1}u_{t-s}^{2}}$, where $u_t$ is the residual of CAPM model\\ 
\bottomrule
\end{tabular}}
{\textit{Note}. This table presents the methods for calculating the standard stock characteristics used in this study. $R_t$ is the return, $V_t$ and $\$V_t$ are share and dollar volume respectively. $askp_t$ and $bidp_t$ are closing ask and bid prices on day $t$, and 
$\mathrm{shrout}_t$ is the shares outstanding.
}
\end{table}

\subsection{Six Back-testing Segments}
We present the six backtesting segments and their corresponding training periods and portfolio components in the Table \ref{tab:backtesting_segments}.
\begin{table}[htbp]
\TABLE
{Backtesting and Training Segments with Portfolio Components\\ \label{tab:backtesting_segments}}
{\begin{tabular}{c|c|c}
\hline
\textbf{Backtesting segment} & \textbf{Training} & \textbf{Portfolio Components} \\
\hline
1993-1997 & 1990-1992 & Top 500 stocks at end of 1992 \\
\hline
1998-2002 & 1995-1997 & Top 500 stocks at end of 1997 \\
\hline
2003-2007 & 2000-2002 & Top 500 stocks at end of 2002 \\
\hline
2008-2012 & 2005-2007 & Top 500 stocks at end of 2007 \\
\hline
2013-2017 & 2010-2012 & Top 500 stocks at end of 2012 \\
\hline
2018-2022 & 2015-2017 & Top 500 stocks at end of 2017 \\
\hline
\end{tabular}}
{\textit{Note}. This table summarizes the period of six backtesting segments with corresponding training periods and portfolio components for the respective periods.}
\end{table}

\end{APPENDICES}
\end{document}